\documentclass[secthm]{elsart}

\usepackage[latin1]{inputenc} 
\usepackage{amsfonts}
\usepackage{amsmath}
\usepackage{amssymb}
\usepackage{latexsym}
\usepackage{graphicx}

\def\nesetril{Ne$\check{\textrm{s}}$et$\check{\textrm{r}}$il~}

\def\hlineny{Hlin$\check{\textrm{e}}$n\'y}

\newcommand{\macro}[2]{\newcommand{#1}[1]{#2}}

\macro{\cwd}{\mathit{cwd}(#1)}
\macro{\twd}{\mathit{twd}(#1)}
\macro{\angl}{\mathop\langle #1 \mathop\rangle}
\macro{\const}{\mathbf{#1}}

\def\Nset{\mathbb{N}}

\def\cA{\mathcal{A}} 
\def\cB{\mathcal{B}}

\def\cF{\mathcal{F}}

\def\cB{\mathcal{B}}

\def\cC{\mathcal{C}}
\def\cT{\mathcal{T}}
\def\cL{\mathcal{L}}

\def\cR{\mathcal{R}}

\def\tG{\widetilde{G}}
\def\tn{\tilde{n}}

\def\up#1{\ulcorner #1\urcorner}
\newcommand{\true}{\textsc{true}}

\newcommand{\floor}[1]{\left\lfloor{#1}\right\rfloor}

\newtheorem{question}{Question}

\newenvironment{exas}[1][Examples]%
               {\par\addvspace{1ex}\noindent{\bf #1}\hspace{0.5em}}%
               { \nopagebreak%
                \strut\nopagebreak %
                \par\addvspace{\medskipamount}
               }

\begin{document}

\begin{frontmatter}

\title{Compact Labelings For Efficient First-Order Model-Checking}


\author{Bruno Courcelle\thanksref{ANR}\thanksref{IUF}} \and
\author{Cyril Gavoille\thanksref{ANR}} \and
\author{Mamadou Moustapha Kant\'e\thanksref{ANR}}

\thanks[ANR]{Supported by the GRAAL project of ``Agence Nationale pour
  la Recherche''.}  \thanks[IUF]{Member of ``Institut Universitaire de
  France''.} 

\address{Universit\'e de Bordeaux, LaBRI, CNRS\\ 351 cours de la
  Libération\\ 33405 Talence, France}


\begin{abstract}
  We consider graph properties that can be checked from labels, i.e.,
  bit sequences, of logarithmic length attached to vertices.  We prove
  that there exists such a labeling for checking a first-order formula
  with free set variables in the graphs of every class that is
  \emph{nicely locally cwd-decomposable}. This notion generalizes that
  of a \emph{nicely locally tree-decomposable} class. The graphs of
  such classes can be covered by graphs of bounded \emph{clique-width}
  with limited overlaps. We also consider such labelings for
  \emph{bounded} first-order formulas on graph classes of
  \emph{bounded expansion}. Some of these results are extended to
  counting queries.
\end{abstract}


\end{frontmatter}

\section{Introduction}\label{sec:1}
The model-checking problem for a class of structures $\cC$ and a
logical language $\cL$ consists in deciding for given $S\in \cC$, and
for some fixed sentence $\varphi \in \cL$ if $S\models \varphi$, i.e.,
if $S$ satisfies the property expressed by $\varphi$. More generally,
if $\varphi$ is a formula with free variables $x_1,\ldots, x_m$ one
may ask whether $S$ satisfies $\varphi(a_1,\ldots,a_m)$ where
$a_1,\ldots,a_m$ are values given to $x_1,\ldots,x_m$. One may also
wish to list the set of $m$-tuples $(a_1,\ldots,a_m)$ that satisfy
$\varphi$ in $S$, or simply count them.

Polynomial time algorithms for these problems (for fixed $\varphi$)
exist for certain classes of structures and certain logical
languages. In this sense graphs of bounded degree ``fit'' with
first-order (FO for short) logic \cite{SEE96,DG07} and graphs of
bounded tree-width or clique-width ``fit'' with monadic second-order
(MSO for short) logic. Frick and Grohe \cite{FRI04,FG01,GRO08} have
defined \emph{Fixed Parameter Tractable} (FPT for short) algorithms
for FO model-checking problems on classes of graphs that may have
unbounded degree and tree-width (definitions and examples are given in
Section \ref{sec:4}) and our results will concern such classes. We
will also use graph classes of \emph{bounded expansion}, a notion
introduced by Ne$\check{\textrm{s}}$et$\check{\textrm{r}}$il and
Ossona de Mendez \cite{NOM06}.

We will use similar tools for the following labeling problem: let be
given a class of graphs $\mathcal{C}$ and a property
$P(x_{1},\ldots,x_{m},Y_{1},\ldots,Y_{q})$ of vertices
$x_{1},\ldots,x_{m}$\ and of sets of vertices $Y_{1},\ldots,Y_{q}$ of
graphs in $\mathcal{C}$. Our aim is to design two algorithms: an
algorithm $\cA$ that attaches to each vertex $x$ of a given graph of
$\cC$ a label $L(x)$, defined as a sequence of $0$'s and $1$'s, and an
algorithm $\cB$ that checks the property
$P(x_{1},\ldots,x_{m},Y_{1},\ldots,Y_{q})$ by using the labels and no
other information about the considered graph. This latter algorithm
must take as input the labels $L(x_{1}),\ldots,L(x_{m})$ and the sets
of labels $L(Y_{1}),\ldots,L(Y_{q})$ of the sets $Y_{1},\ldots,Y_{q}$
and tells whether $P(x_{1},\ldots,x_{m},Y_{1},\ldots,Y_{q})$ is
true. Moreover each label $L(x)$ identifies the vertex $x$ in the
graph. An \emph{$f$-labeling scheme} for a class of structures $\cC$
is a pair $(\cA,\cB)$ of algorithms solving the labeling problem and
using labels of length at most $f(n)$ for $n$-vertex graphs of
$\cC$. Results of this type have been established for monadic
second-order (MSO for short) logic by Courcelle and Vanicat
\cite{CV03} and, for particular properties (connectivity queries, that
are expressible in MSO logic) by Courcelle and Twigg in \cite{CT07}
and by Courcelle et al. in \cite{CGKT08}.

Let us review the motivations for looking for \emph{compact
  labelings}.  By \emph{compact}, we mean of length of order less than 
$O(n)$, where $n$ is the number of vertices of the graph, hence in
particular of length $\log ^{O(1)}(n)$. 

In distributed computing over a communication network with underlying
graph $G$, nodes must act according to their local knowledge
only. This knowledge can be updated by message passing. Due to space
constraints on the local memory of each node, and on the sizes of
messages, a distributed task cannot be performing by representing the
whole graph $G$ in each node or in each message. It must rather
manipulate compact representations of $G$, distributed in a balanced
way over the graph.  For an example, the routing task may use routing
tables that are sub-linear in the size of $G$ (preferably of
poly-logarithmic size), and short addresses transmitted in the headers
of messages (of poly-logarithmic size too). As surveyed in \cite{GP03}
many distributed tasks can be optimized by the use of labels attached
to vertices. Such labels should be usable even when the network has
node or link crashes. They arise from \emph{forbidden-set labeling
  schemes} in \cite{CT07}. In this framework, local informations can
be updated by transmitting to all surviving nodes the list of (short)
labels of all defected nodes and links, so that the surviving nodes
can update their local information, e.g., their routing tables.

Let us comment about using set arguments. The forbidden (or defective)
parts of a network are handled as sets of vertices passed to a query
as an argument. This means that algorithm $\cA$ computes the labels
once and for all, independently of the possible forbidden parts of the
network. In other words the labeling supports node deletions from the
given network. (Edge deletions are supported in the labelings of
\cite{CGKT08} and \cite{CT07}.) 

If the network is augmented with new nodes and links, the labels must
be recomputed. We leave this incremental extension as a topic for
future research.

Set arguments can be used to handle deletions, but also constraints,
or queries like ``what are the nodes that are at distance at most $3$
of $X$ and $Y$'' where $X$ and $Y$ are two specified sets of nodes.

This article is organized as follows. In Section \ref{sec:2} we give
some preliminary definitions regarding first-order logic and we define
the notions of clique-width and of labeling schemes. Section
\ref{sec:3} deals with first-order logic and needed results. In
Section \ref{sec:4} we define the notions of \emph{local bounded
  clique-width} and of \emph{nicely locally cwd-decomposable}. We give
some examples and some preliminary results. Section \ref{sec:5} is
devoted to the proofs of the main results of this article. In Section
\ref{sec:6} we extend some of the main results to counting queries.

\section{Definitions}\label{sec:2}

Our results concern graph properties expressed by logical formulas,
which assumes that graphs are represented by relational
structures. All graphs and relational structures will be finite. 

A \emph{relational signature} is a finite set $\cR = \{R,S,T,
\ldots\}$ of relation symbols, each of which given with an
\emph{arity} $ar(R) \geq 1$. A finite relational $\cR$-structure $S$
is defined as $\angl{D_S,(R_{S})_{R\in \cR}}$ where $R_{S} \subseteq
D_S^{ar(R)}$. The set $D_S$ is called the \emph{domain} of $S$. A
relational signature $\cR$ is binary if $ar(R)\leq 2$ for all $R\in
\cR$. A relational structure is binary if it is a relational
$\cR$-structure for some binary relational signature $\cR$. We let
$\cR_i$ be the set of symbols of arity $i$.

A binary relational $\cR$-structure $S=\angl{D_S,(R_S)_{R\in \cR}}$
will be identified with a colored graph $G$ with vertex set $D_S$,
that has an edge from $x$ to $y$ colored by $R$ in $\cR_2$ if and only
if $R_S(x,y)$ holds, and such that a vertex $x$ has color $P$ in
$\cR_1$ if and only if $P(x)$ holds. Hence, $G$ is a directed graph
such that each edge has a color and a vertex a possibly empty set of
colors\footnote{It is technically useful in many cases to have several
  colors attached to a vertex. Furthermore, these colored graphs
  correspond to relational structures with relation symbols of arity
  $1$ (vertex colors) and $2$ (edge colors)}. We use standard graph
theoretical notations: $V_G$ for vertex set, $E_G$ for edge set and we
will write $G$ as the relational structure $\angl{V_G, (edg_aG)_{a\in
    C_2}, (p_{aG})_{a\in C_1}}$ where $\cR_2=\{edg_a\mid a\in C_2\}$
and $\cR_1=\{p_a\mid a\in C_1\}$. Such a graph is not colored if
$\cR_2=\{edg\}$ and $\cR_1=\emptyset$. A graph represented by the
relational structure $\angl{V_G, (edg_aG)_{a\in C_2}, (p_{aG})_{a\in
    C_1}}$ is called a $C$-graph, $C=C_1\cup C_2$.

Let $G$ be a graph, colored or not. If $X$ is a subset of the set of
vertices of $G$, we let $G[X]$ be the induced sub-graph of $G$ with
vertex set $X$ and induced colors in the obvious way and, we let
$G\backslash X$ be the sub-graph $G[V_G-X]$\footnote{If $X$ is the
  singleton $\{x\}$, we write $G\backslash x$ instead of $G\backslash
  \{x\}$.}. 

If $X$ is a subset of $V_G$, we let $N_G^t(X)$ be the set $\{y\in
V_G\ |\ d(x,y)\leq t\ \textrm{for some}\ x\in X\}$ and $d(x,y)$ is the
length of a shortest undirected path between $x$ and $y$ in $G$.

An undirected graph is a graph $G$ such that $\cR_2=\{edg\}$ and $edg$
is symmetric. If $G$ is any graph and $m\geq 1$, we denote by $G^m$
the simple, loop-free undirected graph such that $V_{G^m}=V_G$ and two
distinct vertices $x$ and $y$ are adjacent in $G^m$ if and only if
$d(x,y)\leq m$.

A graph has \emph{arboricity} at most $k$ if it is the union of $k$
edge-disjoint forests (independently of the colors of its edges and of
its vertices).


We now define \emph{first-order} logic and \emph{monadic second-order}
logic on relational structures and thus, on graphs.  Let $\cR$ be a
relational signature. \emph{Atomic} formulas over relational
$\cR$-structures are $x=y$, $x\in X$ and $R(x_1, \ldots, x_{ar(R)})$
for all relations $R$ in $\cR$. A first order formula (FO formula for
short) over relational $\cR$-structures is a formula formed from
atomic formulas over relational $\cR$-structures with Boolean
connectives $\wedge, \vee, \neg, \Rightarrow$ and first-order
quantifications $\exists x$ and $\forall x$. We may have free set
variables. A monadic second-order formula (MSO formula for short) over
relational $\cR$-structures is formed as FO formulas over relational
$\cR$-structures with set quantifications $\exists X, \forall X$. By
formulas (FO or MSO) we mean formulas written with the signature
appropriate for the considered relational structures.  If the free
variables of a formula $\varphi$ are among $x_1, \ldots, x_m, Y_1,
\ldots, Y_q$ we will write $\varphi(x_1, \ldots, x_m, Y_1, \ldots,
Y_q)$.  A \emph{sentence} is a formula without free variables. We
write $S\models \varphi$ to mean that the sentence $\varphi$ is
satisfied by the relational structure $S$.


The \emph{tree-width} \cite{BOD96} of a graph is independent of edge
directions and of the colors of edges and vertices. It is a well-known
graph parameter and yields many algorithmic properties surveyed by
Grohe \cite{GRO08} and Kreutzer \cite{KRE08}. The survey \cite{BOD07}
by Bodlaender presents tree-width and recent developments about this
notion.

\emph{Clique-width} \cite{CO00} is another graph parameter that yields
interesting algorithmic results. It is sensible to colors and
directions of edges. The original definition of clique-width in
\cite{CO00} concerns only uncolored graphs. However, it can be
extended to colored graphs \cite{CB06,FMR08}.

\begin{defn}[Clique-Width of Colored Graphs] \label{defn:2.1} We let
  $C$ be the finite set ($C=C_1\cup C_2$) of colors for vertices and
  edges. In order to construct graphs, we will use the set
  $[k]:=\{1,2,\ldots, k\}$ for $k\geq 1$ to color also vertices, with
  one and only one color for each vertex. A \emph{$k$-$C$-graph} (or
  \emph{$k$-graph} if $\cR=\{edg\}$) $G$ is defined as
  $G=\angl{V_G,(edg_{aG})_{a\in C_2},(p_{aG})_{a\in C_1},lab_G}$ where
  $lab_G:V_G\to [k]$ is a total function and the other components are
  as defined above. We give several operations on $k$-$C$-graphs. 
  \begin{enumerate}
  \item For $k$-$C$-graphs $G$ and $H$ such that $V_G\cap
    V_H=\emptyset$, we let $G\oplus H$ be the $k$-$C$-graph $K$ where:
    \begin{align*}
      V_K &=V_G\cup V_H,\\ p_{aK}(x)
      &= \begin{cases} p_{aG}(x) & \textrm{if $x\in V_G$}\\ p_{aH}(x)
        & \textrm{if $x\in V_H$} \end{cases}\qquad \textrm{for all
        $a\in C_1$}.\\ edg_{aK}(x,y) &= \begin{cases} edg_{aG}(x,y) &
        \textrm{if $x,y\in V_G$}\\ edg_{aH}(x,y) & \textrm{if $x,y\in
          V_H$} \end{cases}\qquad \textrm{for all $a\in C_2$}.\\
      lab_K(x) &= \begin{cases} lab_G(x) & \textrm{if $x\in V_G$}
        \\ lab_H(x) & \textrm{if $x\in V_H$}.\end{cases}
      \end{align*}
    The graph $G\oplus H$ is well-defined up to isomorphism. 

    \item For a $k$-$C$-graph $G$, for a color $b$ in $C_2$ and for
      distinct $i,j \in [k]$, we denote by $\eta_{i,j}^{b}(G)$, the
      $k$-$C$-graph $K=\angl{V_G,(edg_{aK})_{a\in C_2}, (p_{aG})_{a\in
          C_1}, lab_G}$ where:
    \begin{align*}
      edg_{aK} &= \begin{cases} edg_{aG} & \textrm{if $a\ne b$}\\
      edg_{bG} \cup \{(x,y) ~|~x,y \in V_G~\wedge~x\ne y~\wedge ~i =
      lab_G(x), ~j = lab_G(y)\} & \textrm{if $a=b$}. \end{cases} 
    \end{align*}

  \item For a $k$-$C$-graph $G$, for distinct $i,j \in [k]$, we denote
    by $\rho_{i \to j}(G)$, the $k$-colored graph
    $K=\angl{V_G,(edg_{aG})_{a\in C_2}, (p_{aG})_{a\in C_1}, lab_K}$
    where
    \begin{align*}
      lab_K(x) &= 
      \begin{cases}
        j & \textrm{if $lab_G(x)= i$},\\
        lab_G(x) & \textrm{otherwise}.
      \end{cases}
    \end{align*}

  \item For each $i\in [k]$ and each $A\subseteq C$, $\const{i_A}$
    denotes a $k$-$C$-graph with a single vertex $x$ with
    $lab_{\const{i_A}}(x)=i$ such that $p_{a\const{i_A}}(x)$ holds if
    and only if $a\in A\cap C_1$ and $edg_{a\const{i_A}}(x,x)$ holds
    if and only if $a\in A\cap C_2$. We let
    $C_{C,k}=\{\const{i_A}~|~i\in [k], A\subseteq C\}$.
  \end{enumerate}

  We let $F_{C,k} =\{\oplus, \eta_{i,j}^a,\rho_{i\to j}\mid i,j\in
  [k],a\in C\}$. Each term $t$ in $T(F_{C,k},C_{C,k})$ has a
  \emph{value} $val(t)$: it is the $k$-$C$-graph obtained by
  evaluating $t$ according to definitions (1)-(4). The clique-width of
  a (colored) graph $G$, denoted by $\cwd{G}$, is the minimum $k$ such
  that $G$ is isomorphic to $val(t)$ for some term $t$ in
  $T(F_{C,k},C_{C,k})$.
\end{defn}

There is a function $f:\Nset\times \Nset \to \Nset$ such that if a
$C$-graph has tree-width $w$ then, it has clique-width at most
$f(w,|C|)$. The proof of \cite{CO00} that concerns uncolored graphs
can be adapted. The converse is false because cliques have
clique-width $2$ and unbounded tree-width. For fixed $k$, there exists
a cubic-time algorithm that given an undirected $C$-graph $G$ either
outputs that it has clique-width at least $k+1$ or outputs a term $t$
in $T(F_{C,k'}^a,C_{C,k'}^a)$ that defines $G$ with $k'=2^{k+1}-1$
\cite{OS06,HO07}. This algorithm can be adapted to colored graphs with
$k'=g(k)$ for some function $g$ \cite{KAN08}. Also, every property
expressible in MSO logic can be checked in cubic-time in classes of
colored graphs of bounded clique-width by combining the results of
\cite{CMR00} and of \cite{KAN08}. The survey by Kami\'nski et
al. \cite{KLM08} presents recent results on clique-width.

We now define the notion of \emph{bounded expansion} \cite{NOM06}. As
tree-width, it is independent of colors of vertices and/or
edges. Graph classes with \emph{bounded expansion}, defined in
\cite{NOM06}, have several equivalent characterizations. We will use
the following one.

\begin{defn}[Bounded Expansion] \label{defn:2.2} A class $\cC$ of
  colored graphs has \emph{bounded expansion} if for every integer
  $p$, there exists a constant $N\left(\cC,p\right)$ such that for
  every $G\in \cC$, one can partition its vertex set in at most
  $N\left(\cC,p\right)$ parts such that any $i$ parts for $i\leq p$
  induce a sub-graph of tree-width at most $i-1$.
\end{defn}
The case $i=1$ of Definition \ref{defn:2.2} implies that each part is
a stable set, hence the corresponding partition can be seen as a
\emph{proper vertex-coloring}. We finish these preliminary definitions
by introducing the notion of \emph{labeling scheme}.

\begin{defn}[Labeling Scheme] \label{defn:2.3} Let $\cR$ be a
  relational signature. Let $S=\angl{D_S,(R_G)_{R\in \cR}}$ be a
  relational $\cR$-structure. A \emph{labeling} of $S$ is an injective
  mapping $J:D_S\to \{0,1\}^*$ (or into some more convenient set
  $\Sigma^*$ where $\Sigma$ is a finite alphabet). If $Y$ is a subset
  of $D_S$ we let $J\left(Y\right)$ be the family
  $\left(J\left(y\right)\right)_{y\in Y}$. Clearly each set $Y$ is
  defined from $J\left(Y\right)$.

  Let $\varphi(\bar{x},\overline{Y})$ be an FO or MSO formula over
  relational $\cR$-structures where $\bar{x}$ is an $m$-tuple of FO
  variables and $\overline{Y}$ a $q$-tuple of set variables. Let $\cC$
  be a class of relational $\cR$-structures and let $f:\Nset\to \Nset$
  be an increasing function. An \emph{$f$-labeling scheme supporting
    the query} defined by $\varphi$ in the relational $\cR$-structures
  of $\cC$ is a pair $(\cA,\cB)$ of algorithms doing the following:
  \begin{enumerate}
  \item $\cA$ constructs for each $S$ in $\cC$ a labeling $J$ of $S$
    such that $|J(a)|=O(f(n))$ for every $a\in D_S$, where $n=|D_S|$.
  \item If $J$ is computed from $S$ by $\cA$, then $\cB$ takes as
    input an $(m+q)$-tuple\\ $(J(a_1),\ldots,
    J(a_m),J(W_1),\ldots,J(W_q))$ and says correctly whether:
    \begin{align*}
      S & \models\ \varphi(\bar{a},\overline{W}).
    \end{align*}
  \end{enumerate}
\end{defn}

Labeling schemes based on logical descriptions of queries by MSO
formulas have been first defined by Courcelle and Vanicat \cite{CV03}
for graphs of bounded clique-width (whence also of bounded
tree-width). We recall this theorem. If $\overline{W}$ is a $q$-tuple of
sets, we let $|\overline{W}|=|W_1|+\cdots + |W_q|$ and if $\bar{a}$ is
an $m$-tuple of vertices, we let $|\bar{a}|=m$. 

\begin{thm}\label{thm:2.1} Let $k$ be a positive integer and let $C$
  be a finite set of colors. Then, for every MSO formula $\varphi(x_1,
  \ldots, x_m, Y_1, \ldots, Y_q)$ there exists a $\log$-labeling
  scheme $(\cA,\cB)$ for $\varphi$ on the class of $C$-graphs of
  clique-width at most $k$. Moreover, if the input $C$-graph has $n$
  vertices, algorithm $\cA$ computes the labels $J(x)$ of all vertices
  $x$ in time $O(n^3)$ or in time $O(n\cdot \log(n))$ if the
  clique-width expression of the graph is given. Given $J(a_1),
  \ldots, J(a_m)$ and $J(W_1), \ldots, J(W_q)$ algorithm $\cB$ checks
  whether $\varphi(\bar{a},\overline{W})$ holds in time
  $O(\log(n)\cdot (|\overline{W}|+|\bar{a}|+1))$

  For $n$-vertex $C$-graphs of tree-width at most $k$, algorithm $\cA$
  builds the labelings in time $O(n\cdot \log(n))$.
\end{thm}

The proof of Theorem \ref{thm:2.1} combines the construction of
\cite{CV03} that works for graphs given with their decompositions, and
``parsing'' results by Bodlaender \cite{BOD96} for tree-width and, by
\hlineny, Oum and Seymour \cite{HO07,OS06} and Kanté \cite{KAN08} for
clique-width (discussed above).  Labeling schemes for distance and
connectivity queries in respectively graphs of bounded clique-width
and in planar graphs have been given respectively by Courcelle and
Twigg in \cite{CT07} and by Courcelle, Gavoille, Kant\'e and Twigg in
\cite{CGKT08}.

In the present article, we consider classes of graphs of unbounded
clique-width and, in particular, classes that are \emph{locally
  decomposable} (Frick and Grohe \cite{FRI04,FG01}) and classes of
bounded expansion. So, MSO logic is out of reach for such classes and
we will consider FO logic over $C$-graphs. 

\section{Bounded and Local First-Order Formulas} \label{sec:3}

The definitions below concern binary relational structures called
graphs since they correspond to colored graphs as explained in Section
\ref{sec:2}. Formulas are written over binary relational structures
for a fixed binary relational signature that we do not specify all the
time.

\begin{defn}[Bounded Formulas]\label{defn:3.1} An FO formula
  $\varphi\left(x_1,\ldots,x_m,Y_1,\ldots,Y_q\right)$ is a \emph{basic
    bounded} formula if for some $p\in \Nset$ we have the following
  equivalence for all graphs $G$, all $a_1,\ldots,a_m\in V_G$ and all
  $W_1,\ldots,W_q\subseteq V_G$
  \begin{align*} 
    G \models\ \varphi\left(a_1,\ldots,a_m,W_1,\ldots,W_q\right)~
    \textrm{iff}~ G[X]
    \models\ \varphi\left(a_1,\ldots,a_m,W_1\cap X,\ldots,W_q\cap
    X\right)
  \end{align*}
  for some $X\subseteq V_G$ such that $a_1,\ldots, a_m\in X$ and
  $|X|\leq p$.

  If this is true for $X$, then $G[Y]\models\ \varphi(a_1,\ldots, a_m,
  W_1\cap Y, \ldots, W_q\cap Y)$ for every $Y\supseteq X$. We call $p$
  a \emph{bound on the quantification space}.

  An FO formula is \emph{bounded} if it is a Boolean combination of
  basic bounded formulas. 
\end{defn}

The negation of a basic bounded formula is not (in general) basic
bounded, but it is bounded. The property that a graph has a sub-graph
isomorphic to a fixed graph $H$ is expressible by a bounded formula.

We still call \emph{sentence} an FO formula without free FO variables
that has free set variables.

\begin{defn}[Local Formulas] \label{defn:3.2} An
  FO formula $\varphi\left(x_1,\ldots,x_m, Y_1,\ldots, Y_q\right)$ is
  \emph{$t$-local around $\left(x_1,\ldots,x_m\right)$} if for every
  graph $G$, all $a_1,\ldots,a_m$ in $V_G$ and all subsets
  $W_1,\ldots, W_q$ of $V_G$ we have \begin{align*} G
    \models\ \varphi\left(a_1,\ldots,a_m, W_1,\ldots, W_q\right)~
    \textrm{iff}~ G[N] \models\ \varphi\left(a_1,\ldots,a_m, W_1 \cap
    N, \ldots, W_q\cap N\right) \end{align*} where
  $N=N_G^t(\{a_1,\ldots,a_m\})$.

  An FO sentence $\varphi(Y_1,\ldots, Y_q)$ is \emph{basic
    $(t,s)$-local} if it is equivalent to a sentence of the
  form \begin{align*} \exists x_1\cdots \exists x_s\left(
    \bigwedge_{1\leq i < j \leq s} d\left(x_i,x_j\right) >
    2t\ \wedge\ \bigwedge_{1\leq i\leq s} \psi\left(x_i, Y_1, \ldots,
    Y_q\right)\right) \end{align*} where $\psi\left(x, Y_1, \ldots,
  Y_q\right)$ is $t$-local around its unique free variable $x$.
\end{defn}

\begin{rem} The property $d\left(x,y\right)\leq r$ is basic bounded
  (for $p=r+1$) and $t$-local for $t=\lfloor r/2\rfloor$. Its negation
  $d\left(x,y\right) > r$ is $t$-local and bounded (but not basic
  bounded).
\end{rem}

We now recall a decomposition of $FO$ formulas into $t$-local and
basic $(t',s)$-local formulas due to Gaifman \cite{GAI81}.

\begin{thm}[\cite{LIB04}] \label{thm:3.1} Every  FO formula
  $\varphi(\bar{x}, \overline{Y})$ is logically equivalent to a
  Boolean combination $B\left(\varphi_1(\overline{u_1}, \overline{Y}),
  \ldots, \varphi_p(\overline{u_p},\overline{Y}),
  \right.$$\left.\psi_1(\overline{Y}),\ldots,\psi_h(\overline{Y})\right)$
  where:
  \begin{itemize} 
  \item each $\varphi_i$ is a $t$-local formula around some
    sub-sequence $\overline{u_i}$ of $\bar{x}$,
  \item each $\psi_i$ is a basic $(t',s)$-local sentence.
  \end{itemize}
  Moreover $B$ can be computed effectively and, the integers $t,t'$
  and $s$ can be bounded in terms of $m$ and the quantifier-rank of
  $\varphi$.
\end{thm}

This theorem is usually stated and proved for FO formulas without free
set variables. However, in an FO formula, a set variable $Y_i$ occurs
in atomic formulas of the form ``$y\in Y_i$''. This is equivalent to
``$R_i(y)$'' if $R_i$ is a unary relation representing $Y_i$. We
denote by $\varphi'(\bar{x})$ the formula obtained from
$\varphi(\bar{x}, Y_1, \ldots, Y_q)$ by replacing every sub-formula
``$y\in Y_i$'' by ``$R_i(y)$''. In order to prove that two FO formulas
$\varphi(\bar{x}, Y_1, \ldots, Y_q)$ and $\psi(\bar{x}, Y_1, \ldots,
Y_q)$ are equivalent in every relational structure of a class $\cC$ of
relational $\cR$-structures, it is enough to prove that the
corresponding formulas $\varphi'(\bar{x})$ and $\psi'(\bar{x})$ are
equivalent in every relational structure $S'$ that is an expansion of
a relational structure $S$ in $\cC$ by unary relations $R_1, \ldots,
R_q$.  Hence, Theorem \ref{thm:3.1} follows from its usual formulation
for FO formulas without free set variables. The same holds for Theorem
\ref{lem:3.1} below.

We will use a stronger form of Theorem \ref{thm:3.1} from
\cite{FRI04}, that decomposes $t$-local formulas. Let $m,t\geq 1$. The
\emph{$t$-distance type} of an $m$-tuple $\bar{a}$ is the undirected
graph $\Delta(\bar{a})=([m],edg_{\Delta(\bar{a})})$ where $edg_{\Delta
  (\bar{a})}(i,j)$ iff $d(a_i,a_j)\leq 2t+1$. For each graph $\Delta$
the property that an $m$-tuple $\bar{a}$ satisfies $\Delta(\bar{a})=
\Delta$ can be expressed by a $t$-local formula
$\rho_{t,\Delta}(x_1,\ldots,x_m)$ equivalent to:
\begin{align*}
   \bigwedge_{(i,j)\in edg_{\Delta}} d(x_i,x_j) \leq
   2t+1\ \wedge\ \bigwedge_{(i,j)\notin edg_{\Delta}} d(x_i,x_j) >
   2t+1.
\end{align*}

\begin{thm}[\cite{FRI04}]\label{lem:3.1} Let $\varphi(\bar{x},
  \overline{Y})$ be a $t$-local formula around the $m$-tuple
  $\bar{x}$, $m\geq 1$ with $\overline{Y}=(Y_1,\ldots, Y_q)$. For each
  $t$-distance type $\Delta$ with connected components
  $\Delta_1,\ldots,\Delta_p$ one can compute a Boolean combination
  $F^{t,\Delta}(\varphi_{1,1},\ldots,\varphi_{1,j_1},$\\ $\ldots,
  \varphi_{p,1}, \ldots, \varphi_{p,j_p})$ of~ formulas~
  $\varphi_{i,j}$ with free variables in $\bar{x}$ and in
  $\overline{Y}$ such that:
  \begin{itemize}
  \item The free FO variables of each $\varphi_{i,j}$ belong to
    $\bar{x} \mid\Delta_i$ (where $\bar{x} \mid \Delta_i$ denotes the
    restriction of $\bar{x}$ to $\Delta_i$).
  \item $\varphi_{i,j}$ is $t$-local around $\bar{x} \mid\Delta_i$.
  \item For each $m$-tuple $\bar{a}$, each $q$-tuple of sets
    $\overline{W}$,
    $G\models\ \rho_{t,\Delta}(\bar{a})\ \wedge\ \varphi(\bar{a},
    \overline{W})$ iff $G\models\ \rho_{t,\Delta}
    (\bar{a})\ \wedge\ F^{t,\Delta} ( \ldots,
    \varphi_{i,j}(\bar{a}\mid \Delta_i, \overline{W}), \ldots)$.
  \end{itemize}
\end{thm}

We are interested in on-line checking properties of networks in case
of (reported) failures of some nodes (nodes are vertices of the
associated graphs). Hence, for each property of interest, defined by a
formula $\varphi\left(x_1,\ldots,x_m\right)$, we are not only
interested in checking if
$G\models\ \varphi\left(a_1,\ldots,a_m\right)$ by using
$J\left(a_1\right),\ldots,J\left(a_m\right)$ for $a_1,\ldots,a_m\in
V_G$, but also, in checking if $G\backslash
W\models\ \varphi\left(a_1,\ldots,a_m\right)$ by using
$J\left(a_1\right),\ldots, J\left(a_m\right)$ and $J\left(W\right)$
where $W$ is a subset of $V_G-\{a_1,\ldots,a_m\}$. However, the
property $G\backslash W\models\ \varphi\left(a_1,\ldots,a_m\right)$
for an FO formula $\varphi(x_1,\ldots, x_m)$ is equivalent to $G
\models\ \varphi'\left(a_1,\ldots,a_m,W\right)$ and to
$G_W\models\ \varphi''\left(a_1,\ldots,a_m\right)$ for FO formulas
$\varphi'(x_1,\ldots, x_m,Y)$ and $\varphi''(x_1,\ldots, x_m)$ that
are easy to write. We denote by $G_W$ the graph $G$ equipped with an
additional vertex-color $\perp$, i.e., as the structure $G$ expanded
with a unary relation $p_{\perp}$ such that $p_{\perp G_W}(u)$ holds
iff $u\in W$. We will handle ``holes'' in graphs by means of set
variables.

\section{Locally Decomposable Classes}\label{sec:4}

We will use the same notations as in \cite{FRI04,FG01}. Definition
\ref{defn:4.1} is analogous to \cite[Definition 5.1]{FG01}.

\begin{defn}[Local Clique-Width]\label{defn:4.1}\hfill
  \begin{enumerate} 
  \item The \emph{local clique-width} of a graph $G$ is the
    function $lcw^{G}:\Nset\to \Nset$ defined by $ lcw^{G}(t) :=
    \max\{\cwd{G[N_G^t(a)]}~|~a\in V_G\}$.

  \item A class $\cC$ of graphs has \emph{bounded local
    clique-width} if there is a function $f:\Nset\to \Nset$ such that
    $lcw^G(t) \leq f(t)$ for every $G\in \cC$ and $t\in \Nset$.
  \end{enumerate}
\end{defn}

\begin{exas}[Examples of Graphs of Bounded Local Clique-Width] \hfill
  \begin{enumerate}
  \item Every class of graphs of bounded clique-width has also bounded
    local clique-width since $\cwd{G[A]}\leq \cwd{G}$ for every
    $A\subseteq V_G$ (see \cite{CO00}).
  \item The classes of graphs of bounded local tree-width have bounded
    local clique-width since every class of graphs of bounded
    tree-width has bounded clique-width (see \cite{CO00}). We can cite
    graphs of bounded degree and minor-closed classes of graphs that
    exclude some apex-graph as a minor\footnote{An \emph{apex-graph}
      is a graph $G$ such that $G\backslash u$ is planar for some
      vertex $u$.} (see \cite{FRI04,FG01}) as examples of classes of
    bounded local tree-width.
  \item Let $m$ be a positive integer and let $\cC$ be a class of
    graphs of bounded local clique-width. Then $\cC^m=\{G^m\mid G\in
    \cC\}$ has bounded local clique-width. Let sketch the proof. Let
    $G$ be a graph in $\cC$. For every vertex $x$ of $G$ and every
    positive integer $r$ we have $N_{G^m}^r(x) \subseteq
    N_G^{rm}(x)$. Hence, for every graph $G$ in $\cC$ and for every
    positive integer $r,\ lcw^{G^m}(r) \leq f(rm)$ where $f$ is the
    function that bounds the local clique-width of graphs in $\cC$.

    The same holds for $Line(\cC)=\{Line(G)\mid G\in
    \cC\}$\footnote{If $G$ is a graph we denote by $K=Line(G)$, called
      the \emph{line graph} of $G$, the graph with vertex set the set
      of edges of $G$ and $edg_K(x,y)$ holds if and only if $x$ and
      $y$ are incident.} if $\cC$ has bounded local tree-width. Let
    $G$ be a graph in $\cC$ and let $K=Line(G)$. For every $e$ and
    $e'$ in $E_G=V_K$ we have $d_G(x,y)\leq d_K(e,e')+1$ if $x$ is any
    end vertex of $e$ and $y$ is any end vertex of $e'$. It follows
    that $K[N_K^r(e)]=Line(H)$ where $H$ is a sub-graph of
    $G[N_G^{r+1}(x)]$ and $x$ is an end vertex of $e$. If $\cC$ has
    bounded local tree-width then $\twd{H} \leq \twd{G[N_G^{r+1}(x)}
      \leq f(r)$\footnote{We denote by $\twd{G}$ the tree-width of a
        graph $G$.} for some function $f$, hence $\cwd{K[N_K^r(e)]} =
      \cwd{Line(H)} \leq g(f(r))$ for some function $g$ by a result of
      \cite{GW07}. Hence, the class $\cC$ has bounded local
      clique-width.
  \item The class of interval graphs has not bounded local
    clique-width. Otherwise, interval graphs would have bounded
    clique-width, because if we add to an interval graph a new vertex
    adjacent to all, we obtain an interval graph of diameter $2$.  
  \end{enumerate}
\end{exas}

In order to obtain a $\log$-labeling scheme for certain classes of
graphs of bounded local clique-width, we will cover their graphs, as
in \cite{FRI04,FG01}, by graphs of bounded clique-width. In
\cite{FRI04} a notion of \emph{nicely locally tree-decomposable} class
of structures was introduced. We will define a slightly more general
notion. But before we define the intersection graph of a \emph{cover}
of a graph $G$, i.e., a family $\cT$ of subsets of $V_G$ the union of
which is $V_G$.

\begin{defn}[Intersection Graph] \label{defn:4.2} Let $G$ be a graph
  and let $\cT$ be a cover of $G$. The \emph{intersection graph of
    $\cT$} is the undirected graph $G(\cT)$ where
  $V_{G(\cT)}:=\{x_U~|~U\in \cT\}$ and $x_Ux_V\in E_{G(\cT)}$ if and
  only if $U\cap V\ne \emptyset$.
\end{defn}

\begin{defn}[Graph Covers] \label{defn:4.3} Let $r,\ell\geq 1$ and
  $g:\Nset\to \Nset$. An \emph{$(r,\ell,g)$-cwd cover} of a graph $G$
  is a family $\cT$ of subsets of $V_G$ such that:
  \begin{enumerate}
  \item For every $a\in V_G$ there exists a set $U\in \cT$ such that
    $N_G^r\left(a\right)\subseteq U$.
  \item The graph $G(\cT)$ has degree at most $\ell$. 
  \item For each $U\in \cT$ we have $\cwd{G[U]} \leq g(1)$. 
  \end{enumerate}
  
  An $(r,\ell,g)$-cwd cover is \emph{nice} if condition (3) is replaced by
  condition (3') below:
  \begin{enumerate}
  \item[(3')] For all $U_1,\ldots, U_q\in \cT$ and $q\geq 1$ we have
    $\cwd{G[U_1\cup \cdots \cup U_q]} \leq g(q)$. 
  \end{enumerate}
 
  A class $\cC$ of graphs is \emph{(nicely) locally cwd-decomposable}
  if every graph $G$ in $\cC$ has, for each $r\geq 1$, a (nice)
  $(r,\ell,g)$-cwd cover for some $\ell,g$ depending on $r$ (but not
  on $G$).
\end{defn}

The notions of locally cwd-decomposable and of nicely locally
cwd-decomposable are the same as in \cite{FG01,FRI04} where we
substitute clique-width to tree-width except that our definition
requires nothing about the time necessary to compute covers.

\begin{figure}[!h]
  \centering
  \includegraphics[width=10cm]{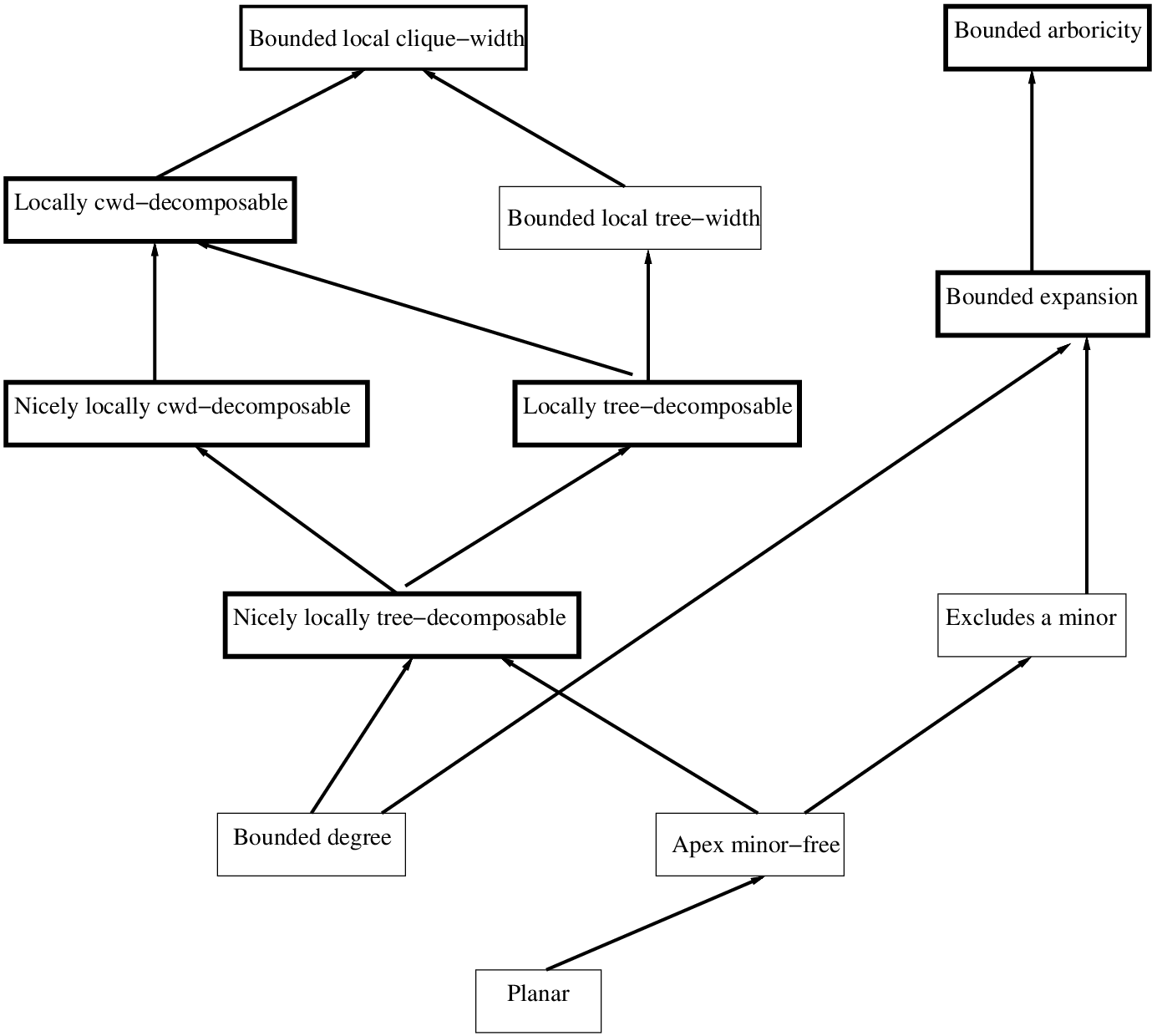}
  \caption{Inclusion diagram indicating which results apply to which
    classes. An arrow means an inclusion of classes. Bold boxes
    are used in this paper.} 
  \label{fig:4.1}
\end{figure}

\begin{exas}[Examples of (Nicely) Locally Cwd-Decomposable Graph Classes]\hfill
  \begin{enumerate}
  \item Every nicely locally cwd-decomposable class is locally
    cwd-decomposable and the converse does not seem to be true (but we
    do not have a counterexample).
  \item Each class of nicely locally tree-decomposable graph is nicely
    locally cwd-decomposable.
  \item We do not know if every graph class of bounded local
    clique-width is locally cwd-decomposable. We conjecture that there
    exists a graph class of bounded local clique-width which is not
    locally cwd-decomposable. 
  \item Figure \ref{fig:4.1} shows inclusion relations between the
    many classes defined in Sections \ref{sec:3} and \ref{sec:4}.
  \end{enumerate}
\end{exas}

\begin{fact}\label{fact:4.1}  The class of unit-interval graphs is
  nicely locally cwd-decomposable.
\end{fact}

\begin{figure}[!h]
  \centering
  \includegraphics[width=0.5\textwidth]{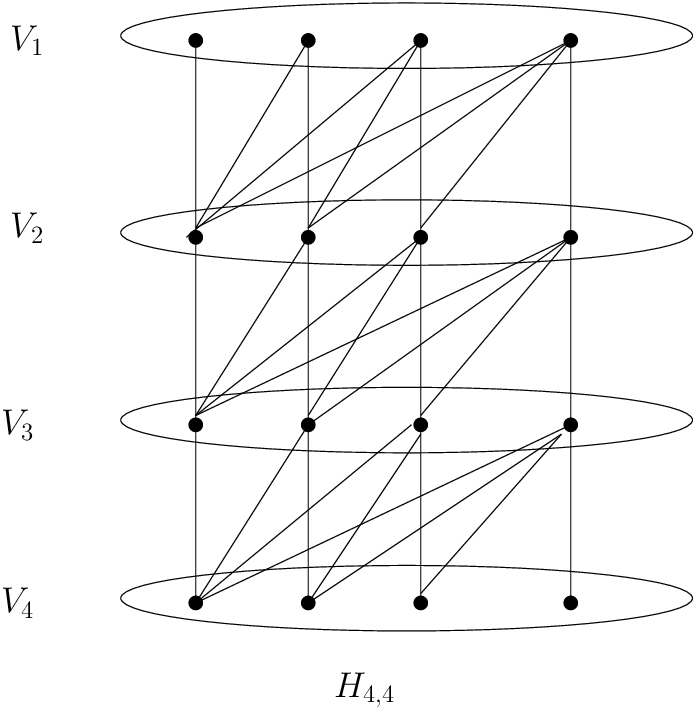}
  \caption[The graph $H_{4,4}$]{The graph $H_{4,4}$. Each $V_i$, for
    $1\leq i \leq 4$, induces a clique.}
  \label{fig:4.2}
\end{figure}

\begin{pf*}{Proof.} We first prove that unit-interval graphs
  have bounded local clique-width. We let $H_{n,m}$ be the graph
  $\angl{V_1\cup \cdots \cup V_n,E_1 \cup E_2}$ with $nm$ vertices
  such that:
  \begin{align*}
    V_i &= \{v_{i,1},\ldots, v_{i,m}\},\\ E_1 &= \bigcup_{1\leq i\leq
      n} \{v_{i,j}v_{i,\ell}\mid j,\ell\leq m\},\\ E_2 &=
    \bigcup_{\overset{1\leq i\leq n-1}{1\leq j\leq m}}
    \{v_{i,j}v_{\ell,j}\mid \ell = i+1,\ldots,m\}
  \end{align*}
  Figure \ref{fig:4.2} shows the graph $H_{4,4}$. Lozin \cite{LOZ07}
  showed that every unit-interval graph with $n$ vertices is an
  induced sub-graph of $H_{n,n}$.

  Let $G$ be a unit-interval graph with $n$ vertices. Then for every
  positive integer $r$ and every vertex $x$ of $G$ the sub-graph
  $G[N_G^r(x)]$ is an induced sub-graph of $H_{r,n}$, i.e., has
  clique-width at most $3r$ since for every positive integers $s$ and
  $t$ the clique-width of $H_{s,t}$ is at most $3s$
  \cite{LOZ07}. (Bagan gives in \cite{BAG08} another proof stating
  that unit-interval graphs have bounded local clique-width.)

  We now prove that the class of unit-interval graphs is nicely
  locally cwd-decomposable. Let $G$ be a unit-interval graph. For
  $1\leq i \leq n-1$ we let $G_i = N_G^{r+1}(v_{i,1})$. It is clear
  that the family $\{G_i\mid 1\leq i \leq n-1\}$ is a nice
  $(r,2r+2,3\cdot (r+1))$-cwd cover of $G$. \qed

\end{pf*}

The lemma below is an easy adaptation of the results in \cite{FG01}.

\begin{lem}\label{lem:4.1} Let $G$ be in a class of graphs of bounded
  local clique-width and let $\varphi$ be a basic $(t,s)$-local
  sentence without set variables. We can check in time $O(n^4)$
  whether $G$ satisfies $\varphi$, $n=|V_G|$.
\end{lem}

\begin{pf*}{Proof Sketch.} Let $G$ be in a class $\cC$ of graphs  of bounded
  local clique-width and let $f$ be the function that bounds the local
  clique-width of graphs in $\cC$. Let $\varphi$ be a basic
  $(t,s)$-local sentence, equivalent to
  \begin{align*}
  \exists x_1\cdots \exists x_s\left( \bigwedge_{1\leq i < j \leq s}
  d\left(x_i,x_j\right) > 2t\ \wedge\ \bigwedge_{1\leq i\leq s}
  \psi(x_i)\right) \end{align*} where $\psi(x)$ is $t$-local around
  its unique free variable $x$. 

  For each vertex $a$ in $G$ we can compute the set $N_G^t(a)$, of
  size at most $n$, in time $O(n^2)$. Since $\cwd{G[N_G^t(a)]} \leq
  f(t)$, we can verify in time $O(n^3)$ if $G$ satisfies $\varphi(a)$
  by combining the results in \cite{HO07} and in \cite{CMR00}. We can
  then compute in time $O(n^4)$ the set $\{a\in V_G\mid G\models
  \varphi(a)\}$. The formula $\varphi$ is valid in $G$ if and only if
  there exist  $a_1, \ldots, a_s$ in $P$ such that $d(a_i,a_j)>2t$. It
  is proved in \cite{GRO08} that we can verify their existence in time
  $O(n^3)$. \qed
\end{pf*}

\section{Labeling Schemes for First-Order Queries} \label{sec:5}

Our results concern $4$ types of graph classes (see Figure
\ref{fig:4.1}) and $5$ types of FO queries. We now state the main
theorem of this section.

\begin{thm}[First Main Theorem]\label{thm:5.1} There
  exist $\log$-labeling schemes $(\cA,\cB)$ for the following queries
  and graph classes. In each case the input graph has $n$ vertices and
  each query is denoted by $\varphi(\bar{x}, \overline{Y})$.
  \begin{enumerate}
  \item Quantifier-free queries in graphs of bounded arboricity.
    Algorithm $\cA$ constructs a labeling in time $O(n)$. Algorithm
    $\cB$ gives the answer in time $O(\log(n)\cdot
    (|\bar{a}|+|\overline{W}|+1))$ for every tuples $\bar{a}$ and
    $\overline{W}$.  The same labeling can be used to check all
    quantifier-free queries.

  \item Bounded FO queries for each class of graphs of bounded
    expansion.  Algorithm $\cA$ constructs a labeling in time
    $O(n)$. Algorithm $\cB$ gives the answer in time $O(\log(n)\cdot
    (|\bar{a}|+|\overline{W}|+1))$ for every tuples $\bar{a}$ and
    $\overline{W}$.

  \item Local queries with set arguments on locally cwd-decomposable
    classes.  Algorithm $\cA$ constructs a labeling in time $O(f(n)+
    n^4)$ where $f$ is the time taken to construct a
    cwd-cover. Algorithm $\cB$ gives the answer in time
    $O(\log(n)\cdot (|\bar{a}|^2+|\overline{W}|+1))$ for every tuples
    $\bar{a}$ and $\overline{W}$.
    
  \item FO queries without set arguments on locally cwd-decomposable
    classes.  Algorithm $\cA$ constructs a labeling in time $O(f(n)+
    n^4)$ where $f$ is the time taken to construct a cwd-cover.
    Algorithm $\cB$ gives the answer in time $O(\log(n)\cdot
    (|\bar{a}|^2))$ for every tuple $\bar{a}$.

  \item FO queries with set arguments on nicely locally
    cwd-decomposable classes.  Algorithm $\cA$ constructs a labeling
    in time $O(f(n)+ n^4)$ where $f$ is the time taken to construct a
    nice cwd-cover.  Algorithm $\cB$ gives the answer in time
    $O(\log(n)\cdot (|\bar{a}|^2+|\overline{W}|+1))$ for every tuples
    $\bar{a}$ and $\overline{W}$.
  \end{enumerate}
\end{thm}


\begin{pf*}{Proof of Theorem \ref{thm:5.1} (1).}
  Let $G$ be a colored graph with $n$ vertices, represented by the
  relational structure $\angl{V_G,(edg_{aG})_{a\in C_2},
    (p_{aG})_{a\in C_1}}$. We recall that $edg_a$ is binary and $p_a$
  is unary.

  Assume that $und(G)$, the graph obtained from $G$ by forgetting edge
  directions and colors of vertices and of edges, is a forest.  Let
  $R$ be a subset of $V_G$ that contains one and only one vertex of
  each connected component, which is a tree, of $G$. For each color
  $a$ in $C_2$ we let $f_a^+,f_a^-:V_G\to V_G$ be mappings such that:
  \begin{enumerate}
   \item[-] $f_a^+(u) = v$\qquad iff\qquad $edg_a(u,v)$ in $G$ and $v$ is on
     the unique undirected \qquad\qquad path  between $u$ and some vertex of $R$
   \item[-] $f_a^-(u) = v$ \qquad iff\qquad $edg_a(v,u)$ in $G$ and
     $v$ is on the unique undirected \qquad\qquad path  between $u$ and
     some vertex of $R$.
  \end{enumerate}
  The edge relation in $G$ is defined by: \begin{align}\label{eq:5.1} edg_{aG}(u,v)
    \Longleftrightarrow\ v=f_a^+(u) \vee \ u=f_a^-(v) \end{align}
  
  If $G$ is the union of $k$ edge-disjoint forests $F_1,\ldots,F_k$ we
  take the pairs $(f_{i,a}^+,f_{i,a}^-)$ for each forest
  $und(F_i)$. The edge relation of $G$ is defined in a similar way as
  in (\ref{eq:5.1}) with $2k$ unary functions by letting
  \begin{align}\label{eq:5.2} edg_{aG}(u,v)
    \Longleftrightarrow\ \bigvee_{i\in [k]} v=f_{i,a}^+(u) \vee
    \ u=f_{i,a}^-(v) \end{align}

  We let $C_1=\{c_1,\ldots, c_{\ell}\}$. For each vertex $x$ of $G$ we
  let $b_x$ be the Boolean vector $(b_{a_1}, \ldots, b_{a_{\ell}})$
  where $b_{a_i}=1$ if and only if $p_{c_iG}(x)$ holds. If vertices
  are numbered from $1$ to $n$ and $\up{x}$ is the bit representation
  of the index of $x$, then we let \begin{align*} J(x) & =
    \big(\up{x}, \up{f_{1,a_1}^+(x)},\up{f_{1,a_1}^-(x)}, \ldots,
    \up{f_{k,a_{\ell}}^+(x)},\up{f_{k,a_{\ell}}^-(x)},
    b_x\big). \end{align*} It is clear that $|J(x)|=O(\log(n))$. We
  now explain how to check any quantifier-free formula.

  Let $\varphi(x_1,\ldots, x_m, Y_1, \ldots, Y_q)$ be a
  quantifier-free formula. For all $m$-tuples $(a_1,\ldots, a_m)$ of
  $V_G$ and all $q$-tuples $(W_1,\ldots, W_q)$ of subsets of $V_G$ we
  can determine $G[\{x_1,\ldots, x_m\}\cup W_1\cup \cdots \cup W_q]$
  from $J(a_1), \ldots, J(a_m)$ and $J(W_1), \ldots, J(W_q)$, and
  check if $\varphi(\bar{a},\overline{W})$ holds.

  It is clear that if the input graph has $n$ vertices and $m$ edges
  then our algorithm constructs the labels in time $O(n+m)$. But, if a
  graph $G$ has arboricity at most $k$, then the number of edges is
  linear in the number of vertices of $G$. Therefore, the labels are
  constructed in linear-time. We now examine the time taken to check
  whether $G$ satisfies $\varphi(a_1, \ldots, a_m, W_1, \ldots,
  W_q)$. For each $x\in \{a_1, \ldots, a_m\}$ it takes constant time
  to check whether $p_{c_iG}(x)$ holds by using the $b_x$ part of
  $J(x)$. For every $x$ and $y$ in $W_1\cup \cdots \cup W_q \cup
  \{a_1, \ldots, a_m\}$ and every $c$ in $C_2$ it takes time
  $O(\log(n))$ to check whether $edg_{cG}(x,y)$ holds and it takes
  time $O(|W_i|\cdot \log(n))$ to check if $x$ is in $W_i$. Therefore
  we can check the validity of $\varphi(a_1, \ldots, a_m, W_1, \ldots,
  W_q)$ in time $O(\log(n) \cdot (|\overline{W}|+|\bar{a}|+1))$ since
  a quantifier-free formula is a Boolean combination of atomic
  formulas.  \qed
\end{pf*}

\begin{pf*}{Proof of Theorem \ref{thm:5.1} (2).} Let $\cC$ be a class
  of graphs of bounded expansion and let $G$ in $\cC$ be a graph with
  $n$ vertices, represented by the relational structure
  $\angl{V_G,(edg_{aG})_{a\in C_2}, (p_{aG})_{a\in C_1}}$. Let
  $\varphi(x_1,\ldots, x_m, Y_1,\ldots, Y_q)$ with $m\geq 1$ be a be a
  basic bounded formula with bound $p$ on the quantification space
  (see Definition \ref{defn:3.1}). We let $N=N(\cC,p)$ and we
  partition $V_G$ into $V_1\uplus V_2\uplus \cdots \uplus V_N$ as in
  the definition (Definition \ref{defn:2.2}) with each $V_i$
  nonempty. (We denote by $\uplus$ the disjoint union of sets.)

  For every $\alpha \subseteq [N]$ of size $p$ we let $V_{\alpha} =
  \bigcup_{i\in \alpha} V_i$ so that the tree-width of $G[V_{\alpha}]$
  is at most $p-1$. Each vertex $x$ belongs to less than $(N-1)^{p-1}$
  sets $V_{\alpha}$. 

  Hence the basic bounded formula $\varphi(\bar{x},\overline{Y})$ is
  true in $G$ iff it is true in some $G[X]$ with $|X|\leq p$, hence in
  some $G[V_{\alpha}]$ such that $x_1,\ldots,x_m\in V_{\alpha}$. For
  each $\alpha$ we construct a labeling $J_{\alpha}$ of
  $G[V_{\alpha}]$ (of tree-width at most $p-1$) supporting query
  $\varphi$ by using Theorem \ref{thm:2.1}. We let $J(x) =\big(
  \up{x}, \{(\up{\alpha},J_{\alpha}(x))~|~x\in V_{\alpha}\}\big)$. We
  have $|J(x)| = O(\log(n))$.

  Given vertices $a_1, \ldots, a_m$ and sets of vertices $W_1, \ldots,
  W_q$ we now explain how to decide the validity of $\varphi(\bar{a},
  \overline{Y})$ by using $J(a_1),\ldots, J(a_m)$ and $J(W_1), \ldots,
  J(W_q)$. From $J(a_1),$ $\ldots,J(a_m)$ we can determine all those
  sets $\alpha$ such that $V_{\alpha}$ contains
  $a_1,\ldots,a_m$. Using the components $J_{\alpha}(\cdot)$ of
  $J(a_1),\ldots$ $,J(a_m)$ and the labels in $J(W_1),\ldots,J(W_q)$
  we can determine if for some $\alpha$,\ $G[V_{\alpha}]
  \models\ \varphi(a_1,\ldots,a_m,W_1\cap V_{\alpha},\ldots,W_q\cap
  V_{\alpha})$ hence whether
  $G\models\ \varphi(a_1,\ldots,a_m,W_1,\ldots,W_q)$.

  It remains to consider the case of a basic bounded formula of the
  form $\varphi(Y_1,\ldots,Y_q)$, i.e., where $m=0$.  For each
  $\alpha$ we determine the truth value $b_{\alpha}$ of
  $\varphi(\emptyset,\ldots,\emptyset)$ in $G[V_{\alpha}]$. The family
  of pairs $(\alpha,b_{\alpha})$ is of fixed size (depending on $p$)
  and is appended to $J(x)$ defined as above (suitably appended as a
  sequence of bits). From $J(W_1),\ldots,J(W_q)$ we get $D =
  \{\alpha~|~ V_{\alpha}\cap (W_1\cup \cdots \cup W_q)\ne
  \emptyset\}$.

  By using the $J_{\alpha}(\cdot)$ components of the labels in
  $J(W_1)\cup \cdots\cup J(W_q)$ we can determine if for some $\alpha
  \in D$ we have $G[V_{\alpha}]\models\ \varphi(W_1\cap
  V_{\alpha},\ldots,W_q\cap V_{\alpha})$. If one is found we can
  conclude positively. Otherwise, we look for some $b_{\beta} = \true$
  such that $\beta\notin D$. The final answer is positive if such
  $\beta$ is found.

  For a Boolean combination of basic bounded formulas
  $\varphi_1,\ldots,\varphi_t$ with associated labelings
  $J_1,\ldots,J_t$ we take the concatenation $J_1(x), J_2(x), \cdots,
  J_t(x)$ of the corresponding labels. It is of size $O(\log(n))$ and
  gives the desired result.

  In \cite{NOM06} \nesetril and Ossona de Mendez described a
  linear-time algorithm that computes the partition $\{V_1, \ldots,
  V_N\}$. The number of sets $V_{\alpha}$ where $\alpha$ is a subset
  of $[N]$ of size $p$ is bounded by $N^p$. Then the number of graphs
  $G[V_{\alpha}]$ is bounded by $N^p$. Then the labeling $J$ is
  constructed in linear-time since each labeling $J_{\alpha}$ is
  constructed in linear-time by Theorem \ref{thm:2.1}.

  We now examine the time taken to check whether $G$ satisfies
  $\varphi(a_1, \ldots, a_m)$. Each vertex $x$ is in less than
  $(N-1)^{p-1}$ sets $V_{\alpha}$. By comparing the sets that contain
  all the $a_i$'s with the sets that contain $a_1$ we can determine in
  time $O(\log(n)\cdot |\bar{a}|)$ the sets $V_{\alpha}$ that contain
  $(a_1, \ldots, a_m)$. For each $V_{\alpha}$ and each $W_i$ we can
  determine in time $O(\log(n)\cdot |W_i|)$ the set $W_i\cap
  V_{\alpha}$. By Theorem \ref{thm:2.1} we can verify in each
  $G[V_{\alpha}]$ in time $O(\log(n)\cdot
  (|\bar{a}|+|\overline{W}|+1))$ whether $G[V_{\alpha}]$ satisfies
  $\varphi(a_1, \ldots, a_m, W_1\cap V_{\alpha}, \ldots, W_q\cap
  V_{\alpha})$ since each $G[V_{\alpha}]$ has bounded
  tree-width. Therefore $\cB$ checks the validity of $\varphi(a_1,
  \ldots, a_m, W_1, \ldots, W_q)$ in time $O(\log(n)\cdot
  (|\bar{a}|+|\overline{W}|+1))$. \qed
\end{pf*}

\begin{pf*}{Proof of Theorem \ref{thm:5.1} (3).}Let $\cC$ be a locally
  cwd-decomposable class of graphs and let $G$ in $\cC$ be a graph
  with $n$ vertices, represented by the structure
  $\angl{V_G,(edg_{aG})_{a\in C_2}, (p_{aG})_{a\in C_1}}$. Let
  $\varphi(\bar{x},Y_1,\ldots,Y_q)$ be a $t$-local formula around
  $\bar{x}=(x_1, \ldots,x_m)$,\ $m\geq 1$. Then
  $G\models\ \varphi(\bar{a},W_1,\ldots,W_q)$ iff
  $G[N_G^t(\bar{a})]\models\ \varphi(\bar{a},W_1 \cap N_G^t(\bar{a}),
  \ldots, W_q\cap N_G^t(\bar{a}))$. Let $\Delta$ be a $t$-distance
  type with connected components $\Delta_1,\ldots, \Delta_p$. By Lemma
  \ref{lem:3.1},
  $G\models\ \rho_{t,\Delta}(\bar{a})\ \wedge\ \varphi(\bar{a},W_1,\ldots,W_q)$
  iff $G\models\ \rho_{t,\Delta}(\bar{a})\ \wedge\ F^{t,\Delta}
  (\varphi_{1,1}(\bar{a} \mid \Delta_1, W_1,\ldots,W_q),
  \ldots,\varphi_{p,j_p}(\bar{a} \mid \Delta_p, W_1,\ldots,W_q))$.

  We let $\cT$ be an $(r,\ell,g)$-cwd cover of $G$ where
  $r=m(2t+1)$. We use this integer $r$ to warranty that if
  $\Delta=\Delta(a_1, \ldots, a_m)$ and $i_1, \ldots, i_k$ in $[m]$
  belong to a connected component of $\Delta$ then, $N_G^t(\{a_{i_1},
  \ldots, a_{i_k}\})\subseteq U$ for some $U$ in $\cT$.  This is so
  because $d_G(a_{i_1}, a_{i_{k'}}) \leq (m-1)\cdot (2t+1)$ for every
  $k'=2, \ldots, k$, hence, if $a\in N_G^t(\{a_{i_1}, \ldots,
  a_{i_k}\})$ we have $d_G(a_{i_1},a)\leq t+(m-1)\cdot (2t+1)\leq
  r$. Hence, $N_G^t(\{a_{i_1}, \ldots, a_{i_k}\})\subseteq
  N_G^r(a_{i_1})\subseteq U$ for some $U$ in $\cT$. For each vertex
  $x$ there exist less than $\ell$ many sets $V$ in $\cT$ such that
  $x\in V$. We assume that each set $U$ in $\cT$ has an index encoded
  as a bit string denoted by $\up{U}$. There are at most $n\cdot \ell$
  sets in $\cT$. Hence $\up{U}$ has length $O(\log(n))$.

  For each set $U$ in $\cT$ we label each vertex in $G[U]$ with a
  label $K_U(x)$ of length $O(\log(n))$ in order to decide if
  $d_{G[U]}(x,y)\leq 2t+1$ or not by using $K_U(x)$ and
  $K_U(y)$\footnote{For checking if $d_G(x,y)\leq 2t+1$, an
    $(r',\ell',g')$-cwd cover suffices, with $r'=2t+1$.} (Theorem
  \ref{thm:2.1}). For each vertex $x$ of $G$ we let
  \begin{align*}
    K(x) = \Big(\up{x},\ \{\big(\up{U},K_U(x)\big)~|~N(x)\subseteq
    U\},\ \{\big(\up{U},K_U(x)\big)~|~N(x)\nsubseteq U\} \Big)
  \end{align*}
  where $N(x) =N_G^{2t+1}(x)$. (We have $x\in N_G^t(x)$ for all $t\in
  \Nset$.) It is clear that $|K(x)|=O(\log(n))$.

  By Theorem \ref{thm:2.1} for each formula $\varphi_{i,j}(\bar{x}\mid
  \Delta_i,Y_1,\ldots,Y_q)$ arising from Theorem \ref{lem:3.1} and
  each $U\in \cT$ we can label each vertex $x\in U$ by some label
  $J_{i,j,U}^{\Delta}(x)$ of length $O(\log(n))$ so that we can decide
  if $\varphi_{i,j}(\bar{a}\mid \Delta_i,W_1,\ldots,W_q)$ holds in
  $G[U]$ by using $\left(J_{i,j,U}^{\Delta}(b)\right)_{b\in
    \bar{a}\ \mid \Delta_i}$ and $J_{i,j,U}^{\Delta}(W_1\cap U),
  \ldots,J_{i,j,U}^{\Delta}(W_q\cap U) $. For each vertex $x$ of $G$
  we let
  \begin{align*} J_{\Delta}(x) :=
    \Big(\big(\up{U},J_{1,1,U}^{\Delta}(x),\ldots,
    J_{1,j_1,U}^{\Delta}(x), \ldots, J_{p,1,U}^{\Delta}(x),\ldots,
    J_{p,j_p,U}^{\Delta}(x)\big)~|~N_G^t(x)\subseteq U\Big).
  \end{align*}
  It is clear that $|J_{\Delta}(x)|=O(\log(n))$ since each $x$ is in
  less than $\ell$ many sets $V$ in $\cT$. There exist at most $k'=
  2^{k(k-1)/2}\ t$-distance type graphs; we enumerate them by
  $\Delta^1,\ldots,\Delta^{k'}$. For each vertex $x$ of $G$ we let
  $J(x) := \left(\up{x},K(x),J_{\Delta^1}(x), \ldots,
  J_{\Delta^{k'}}(x)\right)$. It is clear that $J(x)$ is of length
  $O(\log(n))$.
  
  By hypothesis the cover $\cT$ is computed in time $f(n)$ for $G$ in
  $\cC$ with $n$ vertices. By Theorem \ref{thm:2.1} the labelings
  $K_U$ and $J_{i,j,U}^{\Delta}$ can be constructed in
  cubic-time. Therefore, the labeling $J$ is constructed in time
  $O(f(n)+ n^4)$ since there are less than $\ell\cdot n$ sets $U$ in
  $\cT$.

  We now explain how to decide whether
  $G\models\ \varphi(a_1,\ldots,a_m,W_1,\ldots,W_q)$ by using
  $J(a_1),\ldots, J(a_m)$ and $J(W_1),\ldots,J(W_q)$.
  
  From the labels $K(x)$, we can determine the set
  $\{\up{U}~|~U\in \cT,\ x\in U\}$, hence the family of sets $U\in
  \cT$ such that $W\cap U\ne \emptyset, \ W\subseteq V_G$, where $W$
  is a set argument.

  Since for each vertex $x$ of $G$ there exists a set $U$ in $\cT$
  such that $N_G^r(x)\subseteq U$, for each pair of vertices $(x,y)$
  we have $d_G(x,y)\leq 2t+1$ if and only if $d_{G[U]}(x,y)\leq
  2t+1$. Hence, by using the components $K(a_1),\ldots, K(a_m)$ from
  $J(a_1),\ldots, J(a_m)$ we can construct the $t$-distance type
  $\Delta$ of $(a_1,\ldots, a_m)$; let $\Delta_1,\ldots,
  \Delta_p$ be the connected components of $\Delta$. From each
  $J(a_i)$ we can recover $J_{\Delta}(a_i)$. For each $\bar{a} \mid
  \Delta_i$ there exists at least one $U\in \cT$ such that
  $N_G^t(\bar{a} \mid \Delta_i)\subseteq U$. We can determine these
  sets (there are less than $\ell$ of them) by using the labels in
  $J(b),\ b\in \bar{a} \mid \Delta_i$. We can now decide whether
  $G\models\ F^{t,\Delta}(\varphi_{1,1}(\bar{a} \mid \Delta_1,W_1\cap
  U_1, \ldots, W_q\cap U_1), \ldots,\varphi_{p,j_p}(\bar{a} \mid
  \Delta_p, W_1\cap U_p, \ldots,W_q\cap U_p))$ for some $U_1,\ldots,
  U_p$ determined from $J(a_1),\ldots, J(a_m)$. By using also
  $J(W_1),\ldots, J(W_q)$ we can determine the sets $W_i\cap U_j$ and
  this is sufficient by Theorem \ref{lem:3.1}. 

  We now examine the time taken to check $\varphi(\bar{a},
  \overline{W})$. For each couple $(a_i,a_j)$ it takes time
  $O(\log(n))$ to check if $d(a_i,a_j)\leq 2t+1$. Since there are at
  most $|\bar{a}|^2$ couples, we construct the graph $\Delta$ in time
  $O(\log(n) \cdot |\bar{a}|^2)$. For each connected component
  $\bar{a}\mid \Delta$ we can determine the sets $U$ that contain it
  in time $O(\log(n) \cdot |\bar{a}|)$ (less than $\ell$ such
  sets). By Theorem \ref{thm:2.1} we can check each $\varphi_{i,j}$ in
  time $O(\log(n)\cdot (|\bar{a}|+|\overline{W}|+1))$. Therefore,
  $\cB$ checks the validity of $\varphi(\bar{a}, \overline{W})$ in
  time $O(\log(n)\cdot (|\bar{a}|^2+ |\overline{W}|+1))$.\qed
\end{pf*}

\begin{pf*}{Proof of Theorem \ref{thm:5.1} (4).}  Let $\cC$ be a
  locally cwd-decomposable class of graphs and let $G$ in $\cC$ be a
  graph with $n$ vertices, represented by the structure
  $\angl{V_G,(edg_{aG})_{a\in C_2}, (p_{aG})_{a\in C_1}}$. Let
  $\varphi(x_1,\ldots,x_m)$ be an FO formula without set arguments. By
  Theorem \ref{thm:3.1} the formula $\varphi$ is equivalent to a
  Boolean combination $B(\varphi_1( \bar{x}), \ldots,$ $\varphi_p(
  \bar{x}),\psi_1, \ldots,\psi_h)$ where $\varphi_i$ is a $t$-local
  formula and $\psi_i$ is a basic $(t',s)$-local sentence without set
  variables, for some $t,t',s$.

  By Lemma \ref{lem:4.1} one can decide the validity of each sentence
  $\psi_i$. Let $b = (b_1,\ldots, b_h)$ where $b_i = 1$ if $G$
  satisfies $\psi_i$ and $0$ otherwise. For each $1\leq i\leq p$ we
  construct a labeling $J_i$ supporting query $\varphi_i$ by Theorem
  \ref{thm:5.1} (3) ($G$ belongs to a locally cwd-decomposable class
  and $\varphi_i$ is a $t$-local formula around $\bar{x}$). For each
  vertex $x$ of $G$ we let $J(x) := (\up{x}, J_1(x), \ldots, J_p(x),
  b)$. It is clear that $|J(x)|=O(\log(n))$ since
  $|J_i(x)|=O(\log(n))$. We now explain how to decide whether
  $G\models\ \varphi(a_1,\ldots,a_m)$ by using $J(a_1),\ldots,
  J(a_m)$.

  From $b$ we can recover the truth value of each sentence
  $\psi_i$. By using $J_i(\bar{a})$ we can check if
  $\varphi_i(\bar{a})$ holds. Then, we can check if $B(\varphi_1(
  \bar{x}), \ldots,$ $\varphi_p( \bar{x}),\psi_1, \ldots,\psi_h)$
  holds hence, if $\varphi(\bar{a})$ holds.

  By Lemma \ref{lem:4.1} the validity of each sentence $\psi_i$ is
  checked in time $O(n^4)$. And, by Theorem \ref{thm:5.1} (3), each
  labeling $J_i$ is constructed in time $O(f(n)+ n^4)$ where $f(n)$ is
  the time taken for constructing an $(r,\ell,g)$-cwd cover. Hence,
  the labeling $J$ can be constructed in time $O(f(n)+ n^4)$.  The
  time taken to check the validity of $\varphi(a_1, \ldots, a_m)$ is
  done in time $O(\log(n)\cdot |\bar{a}|^2)$ by Theorem \ref{thm:5.1}
  (3). \qed
\end{pf*}

Before proving Theorem \ref{thm:5.1} (5) we introduce some definitions
and facts.  If $\cT$ is an $(r,\ell,g)$-cwd cover of a graph $G$, then
$G(\cT)$ has maximum degree at most $\ell$. Let $m$ be a positive
integer, a \emph{proper distance-$m$ coloring} of a graph $H$ is a
proper coloring of $H^m$ (see Section \ref{sec:2} for the definition
of $H^m$). Then, in a proper distance-$m$ coloring, vertices at
distance at most $m$ have different colors. A graph $G$ admits a
proper $(d+1)$-coloring if $d$ is its maximum degree. The graph
$G(\cT)$ has maximum degree at most $\ell$, hence, has a proper
distance-$m$ coloring with $\ell^{O(m)}$ colors since $G(\cT)^m$ has
maximum degree at most $\ell\cdot (1 + (\ell-1) + \cdots +
(\ell-1)^{m-1})$.

If $\cT$ is cover of a graph $G$, for each positive integer $t$ and
each set $U$ in $\cT$ we let $K^t(U)$ be the set $\{x\in
U~|~N_G^{t}(x) \subseteq U\}$. We call it the \emph{$t$-kernel} of $U$.

\begin{pf*}{Proof of Theorem \ref{thm:5.1} (5).} Let $\cC$ be a
  nicely locally cwd-decomposable class of graphs and let $G$ in $\cC$
  be a graph with $n$ vertices, represented by the structure
  $\angl{V_G,(edg_{aG})_{a\in C_2}, (p_{aG})_{a\in C_1}}$.  We want a
  labeling for an FO query with set arguments. By Theorems
  \ref{thm:3.1} and \ref{thm:5.1} (3) it is sufficient to define a
  labeling for FO formulas $\varphi(Y_1,\ldots,Y_q)$ of the form:
  \begin{align*}
    \exists x_1\cdots \exists x_m\left( \bigwedge_{1\leq i < j\leq m}
    d(x_i,x_j) > 2t\ \wedge\ \bigwedge_{1\leq i\leq m} \psi (x_i,
    Y_1,\ldots, Y_q) \right) \end{align*} where $\psi(x,Y_1,\ldots,
  Y_q)$ is $t$-local around $x$. We show how to check such formulas by
  means of $\log$-labelings.

  We consider for purpose of clarity the particular case where $m=2$. Let
  $\cT$ be a nice $(r,\ell,g)$-cwd cover of $G$ where $r=2t+1$, and
  let $\gamma$ be a distance-$2$ coloring of $G(\cT)$, the
  intersection graph of $\cT$ (vertices at distance $1$ or $2$ have
  different colors). For every two colors $i$ and $j$ we let $G_{i,j}$
  be the graph induced by the union of the sets $U$ in $\cT$ that are
  colored by $i$ or $j$ (we may have $i=j$).

  \begin{claim}\label{claim:5.1} $\cwd{G_{i,j}} \leq g(2)$. \end{claim}

  \begin{pf*}{Proof of Claim \ref{claim:5.1}.} Let $\cT^2=\{U\cup
    U'\mid U,U'\in \cT,\ U\cap U'\ne \emptyset\}$. The graph $G_{i,j}$
    is a disjoint union of sets in $\cT\cup \cT^2$. This union is
    disjoint because if $U\cup U'$ with $U\cap U'\ne \emptyset$ meets
    some $U''\in \cT$ such that $U''\ne U, \ U''\ne U'$, then we have
    have $\gamma(U) = i,\ \gamma(U') = j\ne i$ and $U''$ meets $U$ or
    $U'$. It can have neither color $i$ nor color $j$ because $\gamma$
    is a distance-$2$ coloring. Since $\cwd{G[U\cup U']} \leq g(2)$,
    we are done because the clique-width of a disjoint union of graphs
    $H_1,\ldots, H_s$ is $\max \{\cwd{H_i}\mid i=1,\ldots, s\}$. \qed
  \end{pf*}

  \begin{claim}\label{claim:5.2} Let $x\in K^{2t}(U)$ and  $y\in
    K^{2t}(U')$ for some sets $U$ and $U'$ in $\cT$. Then $d_G(x,y) >
    2t$ iff $d_{G[U\cup U']}(x,y) > 2t$. \end{claim}

  \begin{pf*}{Proof of Claim \ref{claim:5.2}.} The ``only if direction'' is
    clear since $d_G(x,y) \leq d_{G[U\cup U']} (x,y)$.

    For proving the converse, assume $d_G(x,y) \leq 2t$; there exists
    a path of length at most $2t$ from $x$ to $y$. This path is in
    $U\cup U'$ since $x\in K^{2t}(U)$ and $y\in K^{2t}(U')$. Hence it
    is also in $G[U\cup U']$, hence $d_{G[U\cup U']} \leq 2t$.  \qed
  \end{pf*}

  Let us now give to each vertex $x$ of $G$ the smallest color $i$
  such that $x\in K^{2t}(U)$ and $\gamma(U) = i$. Hence each vertex
  has one and only one color. We express this by $p_i(x)$ where $p_i$
  is a new unary predicate. For each pair $(i,j)$ (possibly $i=j$) we
  consider the formula $\psi_{i,j}$: \begin{align*} \exists x,y \Big(&
    d(x,y) > 2t\ \wedge\ \psi(x,Y_1,\ldots, Y_q)\ \wedge
    \psi(y,Y_1,\ldots,Y_q)\ \wedge\ p_i(x)\ \wedge\ p_j(y)\Big) \end{align*}

  By Theorem \ref{thm:2.1} we can construct a $log$-labeling $J_{i,j}$
  for the formula $\psi_{i,j}$ in the graph $G_{i,j}$. (We recall that
  vertex colors, i.e., additional unary relations, do not increase
  clique-width; the number of relations $p_i$ does not depend on the
  graph $G$.) We compute the truth value $b_{i,j}$ of
  $\psi_{i,j}(\emptyset,\ldots, \emptyset)$ in $G_{i,j}$; we get a
  vector $\vec{b}$ of fixed length. We also label each vertex $x$ by
  its color $\gamma(x)$. We concatenate that $\vec{b}$ and the
  $J_{i,j}(x)$ for $x\in V_{G_{i,j}}$, giving $J(x)$. The coloring
  $\gamma$ uses $O(\ell^2)$ colors. Then, the number of graphs
  $G_{i,j}$ is bounded by $O(\ell^4)$. Therefore $|J(x)|=O(\log(n))$.

  From $J(W_1), \ldots, J(W_q)$ we can determine those $G_{i,j}$ such
  that $V_{G_{i,j}}\cap (W_1 \cup \cdots \cup W_q) \ne \emptyset$, and
  check if for one of them $G_{i,j}\ \models\ \psi_{i,j}(W_1,\ldots,
  W_q)$. If one is found we are done. Otherwise, we use the $b_{i,j}$'s
  to look for $G_{i,j}$ such that $
  G_{i,j}\ \models\ \psi_{i,j}(\emptyset, \ldots, \emptyset)$ {and}
  $(W_1\cup \cdots \cup W_q)\cap V_{G_{i,j}}=\emptyset$.  This gives
  the correct results because of the following facts:

  \begin{itemize}
  \item If $x,y$ satisfy the formula $\varphi$, then $x\in
    K^{2t}(U),\ y\in K^{2t}(U')$ (possibly $U=U'$) and $d_G(x,y) > 2t$
    implies $d_{G_{i,j}}(x,y)>2t$, hence
    $G_{i,j}\ \models\ \psi_{i,j}(W_1,\ldots, W_q)$ where
    $i=\gamma(U)$ and $j=\gamma(U')$.

  \item If $G_{i,j}\ \models\ \psi_{i,j}(W_1,\ldots, W_q)$ then we get
    $G\ \models\ \varphi(W_1,\ldots, W_q)$ by similar argument (in
    particular $d_{G_{i,j}}(x,y)>2t$ implies $d_{G[U\cup U']}(x,y)>2t$
    which implies that $d_G(x,y)>2t$ by Claim \ref{claim:5.2}).
  \end{itemize}
  
  For $m=1$, the proof is similar by using a proper distance-$1$
  coloring $\gamma$ and the graphs $G_{i,i}$ instead of the graphs
  $G_{i,j}$.

  For the case $m> 2$, the proof is the same: one takes for $\gamma$ a
  distance-$m$ proper coloring of the intersection graph, one
  considers graphs $G_{i_1, \ldots, i_m}$ defined as (disjoint) unions
  of sets $U_1\cup \cdots\cup U_m$ for $U_1, \ldots, U_m$ in $\cT$, of
  respective colors $i_1, \ldots, i_m$ and $\cwd{G[U_{1}\cup \cdots
      \cup U_{m}]} \leq g(m)$. 

  By hypothesis, the cover $\cT$ is computed in time $f(n)$ for an
  $n$-vertex graph $G$ in $\cC$. In each graph $G_{i_1,\ldots, i_m}$
  the labeling $J_{i_1,\ldots, i_m}$ is constructed in cubic-time by
  Theorem \ref{thm:2.1}.  The coloring $\gamma$ uses $\ell^{O(m)}$
  colors. Then, the number of graphs $G_{i_1,\ldots, i_m}$ is bounded
  by $\ell^{O(m^2)}$. Hence, the labeling $J$ is computed in time
  $O(f(n)+ n^3)$.

  We now examine the time taken to check the validity of
  $\varphi(\overline{W})$. For each $G_{i_1, \ldots, i_m}$ and each
  $W_i$ it takes time $O(\log(n)\cdot |W_i|)$ to determine $W_i\cap
  V_{G_{i_1, \ldots, i_m}}$. By Theorem \ref{thm:2.1} it takes time
  $O(\log(n)\cdot (|\overline{W}|+1))$ to check in $G_{i_1,\ldots, i_m}$
  the validity of $\varphi(\overline{W})$.  This terminates the proof
  of Theorem \ref{thm:5.1}. \qed
\end{pf*}

Let us ask a very general question: what can be done with labels of
size $O(\log(n))$? Here is a fact that limits the extension of these
results.

Let $\varphi_0(x,y)$ be the $t$-local and bounded FO formula telling
us whether two distinct vertices $x$ and $y$ are connected by a path
of length~$2$:
\begin{align*}
  x\ne y\wedge \exists z \left( z \neq x \land z\neq y \land
  edg(x,z)\right.  \left.\land edg(z,y) \right)
\end{align*}
The adjacency query has a $\log$-labeling scheme for graphs of bounded
arboricity (Theorem \ref{thm:5.1} (1)).

\begin{prop}\label{prop:5.1} Every labeling scheme supporting
  $\varphi_0$ on graphs with $n$ vertices and of arboricity at most
  $2$ requires labels of length at least $\sqrt{\frac{n}{2}}-1$ for some
  graphs.
\end{prop}

\begin{pf*}{Proof.} With every simple, loop-free and undirected graph $G$
  we associate the graph $\tG$ obtained by inserting a vertex $z_{xy}$
  on each edge $xy$.
  \begin{align*}
    V_{\tG} &= V_G \cup \{z_{x,y}\mid x,y\in V_G\ \textrm{and}\ xy\in
    E_G\},\\ 
    E_{\tG} &= \{xz_{x,y}~|~xy\in E_G\}.
  \end{align*}
  The following properties hold.
  \begin{enumerate}
  \item[(1)] $V_G \subseteq V_{\tG}$ and $|V_{\tG}| = |V_G| + |E_G|$.
  \item[(2)] For all $x,y \in V_G$, $xy \in E_G$ if and only if
    $\tG \models \varphi_0(x,y)$.
  \item[(3)] $\tG$ has arboricity at most $2$.
  \end{enumerate}

  The first two points are clear. For the third one we orient each
  edge $e$ of $G$ and we get a directed graph, that we denote by
  $\vec{G}$. We let:
  \begin{align*}
    F_1&=\{xz_{x,y}\mid (x,y)\in E_{\vec{G}}\},\\ F_2&=\{z_{x,y}y\mid
    (x,y)\in E_{\vec{G}}\}. \end{align*} Neither $F_1$ nor $F_2$ has a
  cycle in $\tG$. Then $\tG$ has arboricity at most $2$ since
  $(F_1,F_2)$ is a bipartition of $E_{\tG}$.

  By using a simple counting argument, one can show that every
  labeling scheme supporting adjacency in simple and undirected graphs
  with $n$ vertices requires some labels of size at least $\frac{1}{n}
  \log_2 \left(2^{\binom{n}{2}}\right) = (n-1)/2$ bits. Hence,
  adjacency requires labels of size $\floor{n/2}$ in all graphs. Using
  (2) above, we conclude that any labeling scheme for $\varphi_0$ on
  the graph family $\mathcal{F}_n = \{\tG \mid G\ \textrm{has $n$
    vertices}\}$ requires labels of size at least
  $\floor{\frac{n}{2}}$. Let $\tG$ be in $\cF_n$ and let
  $\tn=|V_{\tG}|$. Using (1) we have $\tn = n + |E_G| \leq
  \frac{n(n+1)}{2}$, i.e., $n\geq \sqrt{2\tn} -1$. Hence, any labeling
  scheme for $\varphi_0$ on $\cF_n$ requires for some graphs with
  $\tn$ vertices labels of size at least $\floor{\frac{\sqrt{2\tn}
        -1}{2}} > \sqrt{\frac{\tn}{2}} - 1$. \qed
\end{pf*}

\section{Extension to Counting Queries} \label{sec:6}

We now consider an extension to counting queries.
 
\begin{defn}[Counting Query]\label{defn:6.1} Let $\varphi(x_1,\ldots,
  x_m, Y_1,\ldots, Y_q)$ be an FO or MSO formula and let $G$ be a
  (colored) graph. For $W_1,\ldots,W_q\subseteq V_G$ we let:
  \begin{align*}
    \#_G\varphi(W_1,\ldots,W_q) & := \Big|\big\{(a_1,\ldots,a_m)\in
    V_G^m~|~G\models\ \varphi(a_1,\ldots, a_m, W_1,\ldots,
    W_q)\big\}\Big|.
  \end{align*}
  
  The \emph{counting query of $\varphi$} consists in determining
  $\#_G\varphi(W_1,\ldots,W_q)$ for given $(W_1,\ldots,W_q)$. If
  $s\geq 2$ the \emph{counting query of $\varphi$ modulo
    $s$} consists in determining $\#_G\varphi(W_1,\ldots,W_q)$ modulo
  $s$ for given $(W_1,\ldots,W_q)$.
\end{defn}

The following theorem is an easy extension of Theorem \ref{thm:2.1}.

\begin{thm} \label{thm:6.1} Let $k$ be a positive integer and, let
  $\varphi(x_1,\ldots, x_m, Y_1, \ldots, Y_q)$ be an MSO formula over
  colored graphs (binary relational structures) and $s\geq 2$. There
  exists a $\log^2$-labeling scheme (resp. a $\log$-labeling scheme)
  $(\cA,\cB)$ on the class of graphs of clique-width at most $k$ for
  the counting query of $\varphi$ (resp. the counting query of
  $\varphi$ modulo $s$). Moreover, if the input graph has $n$ vertices
  then, algorithm $\cA$ constructs the labels in time $O(n^3)$ or in
  $O(n\cdot \log(n))$ if the clique-width expression is given;
  algorithm $\cB$ computes $\#_G\varphi(W_1, \ldots, W_q)$ in time
  $O(\log^2(n)\cdot (|\overline{W}|+1))$ (resp. $O(\log(n)\cdot
  (|\overline{W}|+1))$).
\end{thm}

We will prove a similar theorem for nicely locally cwd-decomposable classes
of graphs and $FO$ formulas. 

\begin{thm}[Second Main Theorem]\label{thm:6.2} Let
  $\varphi(x_1,\ldots, x_m, Y_1, \ldots, Y_q)$ be an $FO$ formula and
  let $s\geq 2$. There exists a $\log^2$-labeling scheme (resp. a
  $\log$-labeling scheme) $(\cA,\cB)$ for the counting query of
  $\varphi$ (resp. the counting query of $\varphi$ modulo $s$) on
  nicely locally cwd-decomposable classes. Moreover, if the input
  graph has $n$ vertices then, algorithm $\cA$ constructs the labels
  in time $O(f(n)+ n^3)$ where $f(n)$ is the time taken to construct a
  nice cwd-cover; algorithm $\cB$ computes $\#_G\varphi(W_1, \ldots,
  W_q)$ in time $O(\log^2(n)\cdot (|\overline{W}|+1))$
  (resp. $O(\log(n)\cdot (|\overline{W}|+1))$).
\end{thm}

We will first prove Theorem \ref{thm:6.2} for particular $t$-local
formulas on locally cwd-decomposable classes.

\begin{defn}[$t$-Connected Formulas] \label{defn:6.2} A formula
  $\varphi(x_1, \ldots, x_m, Y_1, \ldots, Y_q)$ is
  \emph{$t$-connected} if for all $G$, $a_1, \ldots, a_m\in V_G$ and
  $W_1,\ldots, W_q\subseteq V_G$,\ \begin{align*} G
   \models \varphi(a_1,\ldots, a_m, W_1, \ldots, W_q)\quad
  \textrm{iff}\quad  \begin{cases} \bigwedge_{1\leq i < j \leq m}
  d(a_i,a_j)\leq t \ \textrm{and}\\ G[N]
  \models \varphi(a_1,\ldots, a_m, W_1\cap N, \ldots, W_q\cap
  N)\end{cases} \end{align*} where $N=N_G^t(\{a_1, \ldots, a_m\})$.
\end{defn}

\begin{rem}\label{rem:6.1} Let $\varphi(x_1,\ldots, x_m, Y_1,\ldots,
  Y_q)$ be a $t$-connected formula. Then for all $W\supseteq
  N_G^t(a_1, \ldots, a_m)$:
  \begin{align*}
     G &\models\ \varphi(a_1,\ldots, a_m, W_1, \ldots, W_q)\quad
     \textrm{iff}\quad G[W] \models\ \varphi(a_1,\ldots, a_m, W_1\cap
     W, \ldots, W_q\cap W)\\ \intertext{and, since
       $N_G^t(\{a_1,\ldots, a_m\}) \subseteq N_G^{2t}(a_1)$, we have}
     G& \models\ \varphi(a_1,\ldots, a_m, W_1, \ldots, W_q)\quad
     \textrm{iff}\quad G[N_G^{2t}(a_1)] \models\ \varphi(a_1,\ldots,
     a_m, W_1, \ldots, W_q). \end{align*}
\end{rem}

\begin{lem}\label{lem:6.1} Let $\varphi(x_1,\ldots, x_m, Y_1,\ldots,
  Y_q)$ be a $t$-connected formula and let $s\geq 2$. Then, there
  exists a $\log^2$-labeling scheme (resp. a $\log$-labeling scheme)
  $(\cA,\cB)$ for the counting query of $\varphi$ (resp. the counting
  query of $\varphi$ modulo $s$) on locally cwd-decomposable classes
  of graphs. Moreover, if the input graph has $n$ vertices then,
  algorithm $\cA$ constructs the labels in time $O(f(n)+ n^3)$ where
  $f(n)$ is the time taken to construct a cwd-cover; algorithm $\cB$
  computes $\#_G\varphi(W_1, \ldots, W_q)$ in time $O(\log^2(n)\cdot
  (|\overline{W}|+1))$ (resp. $O(\log(n)\cdot (|\overline{W}|+1))$).
\end{lem}

\begin{pf*}{Proof.} Let $\cC$ be a locally cwd-decomposable class of
  graphs and let $\cT$ be a $(2t,\ell,g)$-cwd cover of an $n$-vertex
  graph $G$ from $\cC$. Let $H$ be the intersection graph of $\cT$
  (Definition \ref{defn:4.2}) and let $\gamma$ be a proper coloring of
  $H$ with colors in $[\ell+1]$.

  \begin{claim}\label{claim:6.1} Let $x\in K_G^{2t}(U)$ and $y\in U'$
    with $\gamma(U) = \gamma(U'),\ U\ne U'$. Then $d_G(x,y) > 2t$.
  \end{claim}
  
  \begin{pf*}{Proof of Claim \ref{claim:6.1}.} If this is not the
    case, then $y\in U$ and $x_U$ and $x_{U'}$ are adjacent in $H$.
    This is impossible since they have the same color.\qed
  \end{pf*}
  
  We color each vertex $x$ of $G$ by $i$, the smallest $\gamma(U)$
  such that $x\in K_G^{2t}(U)$. We represent this by the validity of
  $p_i(x)$, as in the proof of Theorem \ref{thm:5.1} (5). For each
  $i\in [\ell+1]$ we let $\varphi_i$ be the formula:
  \begin{align*} \varphi(x_1, \ldots, x_m, Y_1, \ldots,
  Y_q)\ \wedge\ p_i(x_1).\end{align*} Then the following is clear.
    
  \begin{claim}\label{claim:6.2} $\#_G\varphi(Y_1,\ldots,Y_q) =
    \sum\limits_{i}\ \#_G\varphi_i(Y_1,\ldots, Y_q)$.
  \end{claim}
  
  We now show that the counting query of $\varphi$ admits a
  $\log^2$-labeling scheme on $G$. We let
  $V_i=\bigcup\limits_{\gamma(U) = i} \{U~|~U\in \cT\}$.

  \begin{claim} \label{claim:6.3} $\cwd{G[V_i]}\leq g(1)$. \end{claim}

  \begin{pf*}{Proof of Claim \ref{claim:6.3}.} $V_i$ is  a disjoint
    union of sets $U$ from
    $\cT$. From Definition \ref{defn:4.3} each graph $G[U]$ has clique-width
    at most $g(1)$. Therefore $\cwd{G[V_i]}\leq g(1)$. \qed
  \end{pf*}

  \begin{claim}\label{claim:6.4}  $\#_G\varphi_i(Y_1,\ldots, Y_q) =
    \#_{G[V_i]}\varphi_i(Y_1,\ldots, Y_q)$.
  \end{claim}
  
  \begin{pf*}{Proof of Claim \ref{claim:6.4}.} If $\varphi(a_1,\ldots, a_m,
    W_1, \ldots,W_q)$ holds and $p_i(a_1)$ holds then, $a_1\in
    K_G^{2t}(U)$ for some $U$ such that $\gamma(U) = i$. Hence
    $a_2,\ldots, a_m\in N_G^{2t}(a_1)$ and $G[N_G^{2t}(a_1)]
    \models\ \varphi_i(a_1, \ldots, a_m, W_1, \ldots, W_q)$, hence
    $G[V_i] \models\ \varphi_i(a_1, \ldots, a_m,$\\ $W_1, \ldots,
    W_q)$.

    If $G[V_i] \models\ \varphi_i(a_1, \ldots, a_m, W_1, \ldots,
    W_q)$, then $p_i(a_1)$ holds and $d_{G[V_i]}(a_l,a_s) \leq t$ for
    all $l,s\in [m]$. But $d_G(a_l,a_s) = d_{G[V_i]}(a_l,a_s) =
    d_{G[U]}(a_l,a_s)$ where $a_1\in U$ and $\gamma(U) = i$. And since
    $N_G^t(\{a_1, \ldots, a_m\}) \subseteq V_i$ we have $G
    \models\ \varphi_i(a_1, \ldots, a_m,$\\ $W_1, \ldots, W_q)$.  \qed
    \end{pf*}

    By Theorem \ref{thm:6.1} and Claims \ref{claim:6.3} and
    \ref{claim:6.4} there exists a $\log^2$-labeling $J_i$ for the
    counting query of each $\varphi_i$. For each $x\in V_G$ we let
    $J(x) = (J_1(x), \ldots, J_{\ell+1}(x))$. Hence $J$ is a $\log^2$-labeling for the
    counting query of $\varphi$ by Claim \ref{claim:6.2}. By Theorem
    \ref{thm:6.1} labels of size $O(\log(n))$ are sufficient for the
    counting query of each $\varphi_i$ modulo $s$. 

    By Theorem \ref{thm:6.1} each labeling $J_i$ is constructed in
    cubic-time. Therefore, the labeling $J$ is constructed in time
    $O(f(n)+ n^3)$ where $f(n)$ is the time taken for constructing the
    $(2t,\ell,g)$-cwd cover $\cT$ of $G$. By Claim \ref{claim:6.2} and
    Theorem \ref{thm:6.1} $\cB$ computes $\#_G\varphi(W_1, \ldots,
    W_q)$ in time $O(\log^2(n)\cdot (|\overline{W}|+1))$
    (resp. $O(\log(n)\cdot (|\overline{W}|+1))$).\qed
\end{pf*}

We now prove Theorem \ref{thm:6.2}.

\begin{pf*}{Proof of Theorem \ref{thm:6.2}.} Let
  $\varphi(\bar{x},\overline{Y})$ be an $FO$ formula with free
  variables in $\bar{x}=(x_1, \ldots, x_m)$ and in
  $\overline{Y}=(Y_1,\ldots, Y_q)$. By Theorem \ref{thm:3.1} $\varphi$
  is logically equivalent to a Boolean combination of $t$-local
  formulas around $\bar{x}$ and of basic $(t',s)$-local formulas. We
  have proved that each basic $(t',s)$-local formula admits a
  $\log$-labeling scheme on each nicely locally cwd-decomposable class
  of graphs (Theorem \ref{thm:5.1} (5)). It remains to prove that the
  counting query of a $t$-local formula admits a $\log^2$-labeling
  scheme on each nicely locally cwd-decomposable class of graphs
  $\cC$. Let $G$, a graph with $n$ vertices, be in $\cC$

  Let $\psi(\bar{x},Y_1,\ldots, Y_q)$ be a $t$-local formula around
  $\bar{x}=(x_1,\ldots, x_m)$. By Theorem \ref{lem:3.1} we can reduce
  the counting query of $\psi$ to the counting query of finitely many
  formulas of the form $\rho_{t,\Delta} (\bar{x}) \ \wedge
  \ \varphi'(\bar{x}, Y_1,\ldots, Y_q)$ that can be expressed
  as \begin{align*} \varphi'(\bar{x}, Y_1,\ldots, Y_q) :=
    \bigwedge_{1\leq i <j \leq p} d(\bar{x}\mid \Delta_i,\bar{x}\mid
    \Delta_j) > 2t+1\ \wedge\ \bigwedge_{1\leq i \leq p}
    \varphi_i(\bar{x}\mid \Delta_i,Y_1, \ldots,Y_q) \end{align*}
  where each $\varphi_i$ is $t$-local and $(m\cdot
  (2t+1))$-connected. We can assume that $\psi$ is of the form
  $\varphi'(\bar{x}, Y_1,\ldots, Y_q)$.

  Let $\cT$ be a nice $(r,\ell,g)$-cwd cover where $r=m\cdot (2t+1)$
  and let $\gamma$ be a proper distance-$m$ coloring of $G(\cT)$, the
  intersection graph of $\cT$. For every $m$-tuple of colors
  $(i_1,\ldots, i_m)$ we let $G_{i_1,\ldots, i_m}$ be the graph $G[V]$
  where $V$ is the union of all sets $U\in \cT$ such that
  $\gamma(U)\in \{i_1, \ldots,i_m\}$. We have then $\cwd{G[V]}\leq
  g(m)$ (same arguments as in Claim \ref{claim:5.1}). We color each
  vertex with the smallest color $i$ such that $x\in K_G^r(U)$ and
  $\gamma(U)=i$ and we express this by the validity of $p_i(x)$. We
  let $\varphi'_{i_1, \ldots,i_m}$ be
  \begin{align*} \bigwedge_{1\leq i <j \leq p} d(\bar{x}\mid
  \Delta_i,\bar{x}\mid \Delta_j) > 2t+1\ \wedge\ \bigwedge_{1\leq
    \ell \leq p} \left(\varphi_{\ell}(\bar{x}\mid \Delta_{\ell},Y_1,
  \ldots,Y_q)\ \wedge\ p_{i_{\ell}}(z_{\ell})\right)
  \end{align*} where $z_{\ell}$ is the first variable of each tuple
  $\bar{x}\mid \Delta_{\ell}$. We have:

  \begin{claim}\label{claim:6.5}  $\#_G\psi(Y_1,\ldots,Y_q) = 
    \sum\limits_{(i_1,\ldots,i_m)}\#_G\varphi'_{i_1,\ldots,
      i_m}(Y_1,\ldots, Y_m)$.
  \end{claim}

  We let $H=G_{i_1,\ldots,i_m}$.  By the same arguments as in the
  proof of Claim \ref{claim:5.2} we have:

  \begin{claim}\label{claim:6.6} $d_G(\bar{x}\mid
    \Delta_i,\bar{x}\mid \Delta_j) >2t+1$ if and only if
    $d_H(\bar{x}\mid \Delta_i,\bar{x}\mid \Delta_j) >2t+1$.
  \end{claim}

  It follows that:

  \begin{claim}\label{claim:6.7} $\#_G\varphi'_{i_1,\ldots, i_m}(Y_1,
    \ldots, Y_q) = \#_H\varphi'_{i_1,\ldots,i_m}(Y_1,\ldots,
    Y_q)$. 
  \end{claim}

  By Theorem \ref{thm:6.1} and Claims \ref{claim:6.5}, \ref{claim:6.6}
  and \ref{claim:6.7} there exists a $\log^2$-labeling scheme for the
  counting query of each $t$-local formula, and a $\log$-labeling
  scheme is enough for modulo counting.

  By hypothesis, a nice $(r,\ell,g)$-cwd cover $\cT$ of $G$ can be
  constructed in time $f(n)$. For each formula $\varphi_{i_1,\ldots,
    i_m}$ the associated labeling $J_{i_1,\ldots, i_m}$ is constructed
  in time $O(n^3)$ by Theorem \ref{thm:6.1}. The coloring $\gamma$
  uses $\ell^{O(m)}$ colors. The number of graphs $G_{i_1,\ldots,
    i_m}$ is bounded by $\ell^{O(m^2)}$. Hence, the labeling $J$ is
  computed in time $O(f(n)+ n^3)$. By Claim \ref{claim:6.5} and
  Theorem \ref{thm:6.1} algorithm $\cB$ computes $\#_G\varphi(W_1,
  \ldots, W_q)$ in time $O(\log^2(n)\cdot (|\overline{W}|+1))$
  (resp. $O(\log(n)\cdot (|\overline{W}|+1))$). This finishes the
  proof.\qed
\end{pf*}

\section{Conclusion}

We conjecture that the results of Theorem \ref{thm:5.1} (3-5) extend
to classes of graphs that exclude, or locally exclude a minor
(definitions are from \cite{DGK07,GRO08}).

\begin{question}\label{quest:1} Does there exist a $\log$-labeling
  scheme for FO formulas with set arguments on locally
  cwd-decomposable classes?
\end{question}



\end{document}